\documentclass[prc]{revtex4}

\usepackage{epsfig}
\usepackage{graphicx}
\usepackage[bookmarks,dvips,pdfhighlight=/O,pdfstartview=FitH]{hyperref}

\newcommand{\GeV}{\; \mathrm{GeV}}

\newcommand{\pq}{{q\cdot p}}

\begin{document}

\title{Quark--hadron duality in lepton scattering off nuclei}
\author{Olga Lalakulich}
\altaffiliation[Current affiliation: Institute for Theoretical Physics, Giessen University, Germany]
\email[Contact e-mail: Olga.Lalakulich@theo.physik.uni-giessen.de]
\author{Natalie Jachowicz}
\author{Christophe Praet}
\author{Jan Ryckebusch}
\affiliation{Department of Subatomic and Radiation Physics, Ghent University, Belgium}

\begin{abstract}
A phenomenological study of quark--hadron duality in electron and
neutrino scattering on nuclei is performed. We compute the structure
functions $F_2$ and $xF_3$ in the resonance region within a framework
that includes the Dortmund-group model for the production of the
{f}{i}rst four lowest-lying baryonic resonances and a relativistic
mean-field model for nuclei. We consider four-momentum transfers
between 0.2 and 2.5 GeV$^2$. The results indicate that nuclear effects
play a different role in the resonance and DIS region. We find that
global but not local duality works well. In the studied range of
four-momentum transfers, the integrated strength of the computed
nuclear structure functions in the resonance region, is considerably
lower than the DIS one.
\end{abstract}

\maketitle

\section{Introduction}

Nearly forty years ago, Bloom and Gilman found \cite{Bloom:1970xb} that in electron
scattering on protons the inclusive structure function $F_2$ in the
resonance region oscillates around the DIS scaling curve and, after
averaging, closely resembles it. This phenomenon is one of the ways
quark--hadron duality reveals itself in physical processes. Generally
quark--hadron duality establishes a relationship between the
quark--gluon description of a certain phenomenon, which is
theoretically justified in the DIS region, and the hadronic
description, which is more convenient at medium and low energies.
Understanding duality is also essential when establishing
relationships between exclusive and inclusive processes. For a recent
and detailed review of duality we refer the reader to
Ref.~\cite{Melnitchouk:2005zr}.

So far, most theoretical studies of quark--hadron duality in lepton
scattering were dealing with nucleon targets. The topic becomes of
great practical interest when turning to nuclear targets and neutrino
sources.  The current precision measurements of the oscillation
parameters require an efficient and accurate description of the
neutrino--nucleus cross sections. Of particular interest is the
resonance region and the possibility of linking it with the DIS
region. A hadronic description of the neutrino-nucleus cross sections
at low $Q^2$ requires the vector and axial transition form factors for
each resonance. For the majority of the resonances, these transition
form factors are not well constrained. Provided that one can establish
that quark-hadron duality holds with a reasonable accuracy, one could
think of using the DIS results for estimating the neutrino-nucleus
cross sections in the resonance region. In that respect it is worth
mentioning that in nuclei, the Fermi motion of the nucleons smears the
observables, so that the averaging in the resonance region required
for duality, proceeds to a certain extent automatically.  The issue
whether quark--hadron duality holds with sufficient accuracy in
lepton-nucleus scattering, requires further theoretical and
experimental investigation. The present paper addresses this issue
from the theoretical point of view.

Recent electron scattering measurements at Jefferson Laboratory (JLab)
have confirmed the validity of Bloom--Gilman duality for the proton,
deuterium \cite{Niculescu:2000tk} and iron \cite{Arrington:2003nt}
structure functions. Further experimental efforts are required for
neutrino scattering. Among the upcoming neutrino experiments,
Miner$\nu$a\cite{Boehnlein:2007zz,SolanoSalinas:2007zza,minerva} and
SciBooNE\cite{AlcarazAunion:2007zz,Hiraide:2007zz,sciboone} aim at
measurements with carbon, iron and lead nuclei as targets. From the
theoretical side, recent investigations of the phenomenon of duality
for electron and neutrino scattering on nucleons include the works
reported in Refs.~
\cite{Matsui:2005ns,Graczyk:2005uv,Lalakulich:2006yn}.  These studies
differ in the way they treat the resonant contributions and the way
they parameterize the DIS structure functions. This paper extends the
study of Ref.~\cite{Lalakulich:2006yn} about the duality phenomenon in
the nucleon to nuclei.

For a free nucleon target, the structure functions generally depend on
the transferred energy $\nu=E-E'$ and four-momentum $Q^2=-q_\mu
q^\mu$. At low $Q^2$ the $\nu$--distributions reveal several peaks,
which correspond to various baryon resonances.  We briefly sketch our
theoretical approach to resonance production in nuclei in
Section~\ref{nucleus}. The nuclear structure functions are defined in
Section~\ref{defSF}. At high $Q^2$ the structure functions exhibit
scaling behavior, which is discussed in Section~\ref{DIS}. Comparing
the structure functions in these two regions allows one to check the
basic features of duality and compare its validity for different
targets and incoming leptons. Our results are presented in
Section~\ref{electrons} for electrons and Section~\ref{neutrinos} for
neutrinos. Conclusions are given in Section~\ref{summary}.
 

\section{Formalism}

We consider inclusive charged-current (CC) neutrino scattering from nuclei and its electromagnetic counterpart
\begin{equation}
\nu_l(k^\mu) + A \to l^- (k^{\prime \mu}) + X \ ,
\qquad 
l^- (k^\mu) + A \to l^- (k^{\prime \mu}) + X \ ,
\end{equation}
where $l$ is the lepton flavor, $A$ represents a nucleus with mass
number $A$, and $k^\mu=(E,\vec k)$ and $k^{\prime\mu}=(E',\vec k')$ are
the four--momenta of the incoming and outgoing lepton respectively. We work in the
laboratory frame of reference. The coordinate system is chosen such
that the $z$-axis lies along the direction of the virtual photon, so
that the transferred momentum is given by
$q^\mu=k^\mu-k^{\prime\mu}=(\nu, 0,0, q^z)$. The lepton scattering
proceeds in the $xz$--plane. In this section, we investigate the structure functions
$F_2$, $2xF_1$ and $xF_3$, the latter being nonzero for neutrino
reactions only. To this end, CP-violation effects are neglected for
the case of electron scattering. 

\subsection{Resonance production on a nucleus \label{nucleus}}
 
\begin{figure}[hbt]
\epsfig{figure=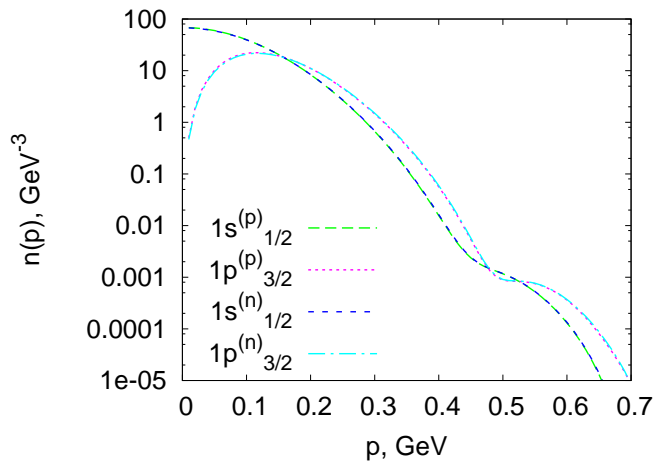,angle=-90,width=0.5\textwidth}
\caption{Momentum distributions for proton and neutron shells in carbon.}
\label{fig:n_alpha}
\end{figure}

For lepton--nucleus scattering we describe the struck nucleus as a
collection of bound nucleons. Assuming an independent--particle shell
model, each nucleon occupies a nuclear shell $\alpha$ with a
characteristic binding energy $e_{\alpha}$ and is described by the
bound--state spinor $u_\alpha$. In the impulse approximation, an
impinging lepton interacts with a single bound nucleon. Hence, the
nuclear cross section can be expressed as an incoherent sum over all
nucleons of one--nucleon cross sections weighted with the
corresponding nucleon momentum distributions $n_\alpha$. For example,
for a carbon nucleus, one has
\[
\begin{array}{l} \displaystyle 
\frac{  d\sigma^{ {}_{\; 6}^{12}C }  }{dQ^2 d\nu} = \int d^3 p   \biggl[ 
2 \frac{d\sigma_{\nu p}   \left|_{1s^{1/2}} \right. }{dQ^2 d\nu}  n^{(p)}_{1s^{1/2}}(|\vec p|)
+ 4 \frac{d\sigma_{\nu p}  \left|_{1p^{3/2}} \right. }{dQ^2 d\nu} n^{(p)}_{1p^{3/2}}(|\vec p|)
\\[3mm] \displaystyle   \hspace*{27mm}  \displaystyle 
+ 2 \frac{d\sigma_{\nu n} \left|_{1s^{1/2}} \right.}{dQ^2 d\nu}   n^{(n)}_{1s^{1/2}}(|\vec p|)
+ 4 \frac{d\sigma_{\nu n}  \left|_{1p^{3/2}} \right.}{dQ^2 d\nu}  n^{(n)}_{1p^{3/2}}(|\vec p|)
\biggr]. 
\end{array}
\]
This allows us to employ the one--body lepton-nucleon vertex that can
be well constrained in experiments with a proton and deuteron target.
The four--momentum of the bound nucleon 
can be written as $p^\mu=(m_N-e_\alpha, \vec{p})$. Both the
bound--state spinor $u_\alpha(\vec p)$ and the corresponding binding
energies are computed in the Hartree approximation to the
$\sigma-\omega$ Walecka--Serot
model~\cite{Serot:1984ey,Furnstahl:1996wv}. Binding energies for
carbon and iron are summarized in Table~\ref{tab:e0}.  For each shell,
the nucleon momentum distribution $n_\alpha(|\vec p|)$ is constructed
from the bound--state spinors, the normalization convention being
\[
    \int d^3p \;  n_\alpha(|\vec p|) = 1 \; .
\]
These $n_\alpha(|\vec p|)$ are shown in Fig.~\ref{fig:n_alpha}, for
the case of a carbon nucleus. Clearly, for a specific shell, the proton and
neutron distributions are almost identical.   

\begin{table}
\caption{Binding energies (MeV) for carbon and iron nuclei}
\[
 \begin{array}{ccc}
& proton & neutron
\\
{}^{12}C: & &
\\
1s^{1/2} & 47.76 & 51.17
\\
1p^{3/2} & 16.76 & 19.87
\\[2mm]
{}^{56}Fe: & &
\\
1s^{1/2} & 57.19 & 63.66
\\
1p^{3/2} & 43.11 & 50.12
\\
1p^{1/2} & 39.32 & 46.00
\\
1d^{5/2} & 27.64 & 34.84
\\
2s^{1/2} & 17.77 & 24.41
\\
1d^{3/2} & 16.55 & 23.01
\\
1f^{7/2} & 12.11 & 19.17
\\
2p^{3/2} & -     & 5.99
\end{array}
\]
\label{tab:e0}
\end{table}

After the interaction takes place inside the nucleus, the reaction
products can escape the nucleus without interactions or they can
undergo elastic and/or inelastic rescatterings with the other
nucleons. Thus, the reaction strength is redistributed between
different channels. All these processes are called the final state
interactions (FSI). The effect of FSI can be large for a specific
exclusive process, for example for quasi-elastic nucleon knockout
\cite{Martinez:2005xe}, where the cross section can be suppressed by a
factor of 2. In one--pion production, the outgoing pion can be
absorbed in the nucleus and thus mimic a quasi--elastic event. 
For a duality study, however, it suffices to consider inclusive
reactions. Consequently, since the outgoing hadrons and the residual
nucleus are not detected, we can make the assumption, following
Ref.~\cite{Benhar:2006nr}, that FSI can be disregarded.

Recently, duality in lepton--nucleon scattering was investigated
theoretically within the Sato--Lee \cite{Sato:2003rq}, Rein--Sehgal
\cite{Rein:1980wg} and Dortmund--group \cite{Lalakulich:2006sw}
models for resonance production. In this paper, we follow the approach
used in \cite{Lalakulich:2006sw} and extend it to calculate the
nuclear structure functions. In particular, in the resonance region we
take into account the first four low--mass baryon resonances
$P_{33}(1232)$, $P_{11}(1440)$, $D_{13}(1520)$, $S_{11}(1535)$ and
describe the vertices of their leptoproduction within a
phenomenological form-factor approach.
The nucleon structure functions ${\cal W}_i$ are defined by the standard expansion of the hadronic tensor
\begin{equation}
 W_{\mu\nu}=
-g_{\mu\nu}{\cal W}_1 
+  \frac{p_\mu p_\nu}{m_N^2} {\cal W}_2 
- i\varepsilon_{\mu\nu\lambda\sigma} \frac{p^\lambda q^\sigma}{2 m_N^2} {\cal W}_3 
+  \frac{q_\mu q_\nu}{m_N^2} {\cal W}_4 
+  \frac{p_\mu q_\nu + p_\nu q_\mu}{m_N^2} {\cal W}_5 \ .
\label{Wmunu-nucleon}
\end{equation}
Each ${\cal W}_i$ depends on two independent kinematic variables, for
example $Q^2$ and $\nu$, which are determined exclusively by the
lepton kinematics.  Another set of variables, namely $Q^2$ and $W$, is
also possible, since the invariant mass $W$, defined as $W^2=(p+q)^2$,
for a free target nucleon can be uniquely related to $Q^2$ and $\nu$:
$W^2=m_N^2 +2m_N\nu -Q^2$. The analytical expressions for the
one--nucleon structure functions $F_1=m_N {\cal W}_1$, $F_2=\nu {\cal
W}_2$, $F_3=\nu {\cal W}_3$ in terms of form factors for a free
nucleon as well as the form factors themselves are given in
\cite{Lalakulich:2006sw}.  The Fermi motion of the bound nucleon
modifies the expression for the scalar product $(\pq)$, so that the
invariant mass $W^2=(p+q)^2$ will now depend on the nucleon momentum
and binding energy. The variables $Q^2$ and $\nu$, being determined by
lepton kinematics only, remain unaffected. Strictly speaking, the
expansion in Eq.~(\ref{Wmunu-nucleon}) is only valid for a free
(on--mass shell) target nucleon. For a bound nucleon, all inclusive
observables depend not only on $\nu$ and $Q^2$, but also on an
additional independent kinematical variable, which can be chosen to be
$p_\mu  p^\mu=p^2$. Here, we make the assumption (see
\cite{Ferree:1995fb} for a detailed discussion) that expression
(\ref{Wmunu-nucleon}) can still be used to define the bound--nucleon
structure functions, and recalculate them keeping the kinematical
variable $p^2$ as an independent one. The results are given below for
the $W_2$ and $W_3$ structure functions. Equating $p^2=m_N^2$, the
free-nucleon results of \cite{Lalakulich:2006sw} are easily
reproduced.  For the spin-3/2 resonances ($P_{33}(1232)$ and
$D_{13}(1520)$ in our case) one has
\begin{equation}
{\cal W}_i(Q^2,\nu, p^2)=\frac{2}{3m_N} V_i(Q^2,\nu,p^2) R(W,M_R),
\end{equation}
where $R(W,M_R)$ is the finite representation of the $\delta-$function
$\delta(W^2 - M_R^2)$, which gives the relativistic Breit--Wigner distribution:
\[
 R(W,M_R)=\frac{M_R \Gamma_R}{\pi} \frac{1}{(W^2-M^2_R)^2+M_R^2 \Gamma_R^2} \ ,
\]
and the $V_i$ are given below. The upper and lower signs are for the positive ($P_{33}(1232)$) and negative ($D_{13}(1520)$) parity resonances, respectively.

\begin{eqnarray} 
V_2 &=& 
\frac{(C_3^V)^2 + (C_3^A)^2}{M_R^2} Q^2 \left[ \pq +p^2 +M_R^2 \right]  
+\left( \frac{(C_4^V)^2}{m_N^2} 
          + \frac{(C_5^V)^2(Q^2+M_R^2)}{m_N^2 M_R^2} 
          +\frac{2 C_4^V C_5^V}{m_N^2}                   \right)Q^2 
   \left[ \pq +p^2 \mp m_N M_R \right]  
\nonumber
\\[3mm] 
&+&\frac{C_3^V C_4^V}{m_N M_R}  Q^2 \left[ \pq + p^2 + M_R^2 \mp 2 m_N M_R \right]  
+\frac{C_3^A C_4^A}{m_N M_R}  Q^2 \left[ \pq + p^2 + M_R^2 \pm 2 m_N M_R \right]  
+ C_3^A C_5^A \frac{m_N}{M_R} Q^2
\nonumber
\\[3mm] 
&+&\frac{C_3^V  C_5^V}{m_N M_R } Q^2 \left[ \pq + p^2 + M_R^2 \mp 2 m_N M_R +Q^2\right]
+\left[ ({C_5^A})^2 \frac{m_N^2}{M_R^2} 
+ \frac{(C_4^A)^2}{m_N^2} Q^2 \right]
\left[  \pq+ p^2 \pm m_N M_R \right],
\label{calW2}
\end{eqnarray}

\begin{eqnarray} 
V_3 &=& 
2\frac{C_3^V C_3^A}{M_R^2} \left[ 2(Q^2-\pq)^2 +M_R^2(3Q^2-4\pq) \right]
+2\left[ \frac{C_4^V C_4^A}{m_N^2}(Q^2 -\pq) -C_4^V C_5^A \right] (Q^2-\pq)
\nonumber 
\\[3mm] 
&+& 2\frac{C_5^V C_3^A \pq - C_4^V C_3^A(Q^2-\pq)}{M_R m_N} 
\left[2M_R^2 \mp 2 m_N M_R +Q^2 -\pq \right] 
+ 2\left[ C_5^V C_5^A -\frac{C_5^V C_4^A}{m_N^2}(Q^2-\pq) \right] \pq
\nonumber 
\\[3mm] 
& + & 2 \left[C_3^V C_5^A \frac{m_N}{M_R} - 
   \frac{C_3^V\, C_4^A}{M_R m_N} (Q^2-\pq)
  \right]
\left(2 M_R^2 \pm 2 m_N M_R + Q^2 - \pq \right)
\label{calW3}
\end{eqnarray}

For spin-1/2 resonances we have 
\[
{\cal W}_i(Q^2,\nu, p^2)=\frac{1}{m_N} V_i(Q^2,\nu, p^2) R(W,M_R) \; ,
\]
where 
\begin{equation}
V_2=2 m_N^2 \left[\frac{(g_1^V)^2}{\mu^4}Q^4  + \frac{(g_2^V)^2}{\mu^2}Q^2 + (g_1^A)^2 \right] \ ,
\end{equation}
\begin{equation}
V_3= 4 m_N^2 \left[   \frac{g_1^V g_1^A}{\mu^2}Q^2  + \frac{g_2^V g_1^A}{\mu}(M_R\pm m_N)  \right] \ ,
\end{equation}
and $\mu=m_N+M_R$. The upper and lower signs again correspond to
positive ($P_{11}(1440)$) and negative ($S_{11}(1535)$) parity
resonances, respectively. In the case of electroproduction, all axial
form factors should be put equal to zero and the weak vector form
factors should be replaced by the electromagnetic ones for proton or
neutron, depending on the target nucleon.
To make the article self-contained, we present the transition form
factors for each resonance. Electromagnetic and weak vector form factors were determined in 
\cite{Lalakulich:2006sw} by fitting the electroproduction data on
helicity amplitudes in the region $Q^2<3\GeV^2$. Recently, it was
shown \cite{Vereshkov:2007xy} that in order to satisfy the asymptotics
for helicity amplitudes at $Q^2\to \infty$, as prescribed by
perturbative QCD, the vector form factors should also exhibit a
certain asymptotic $Q^2$ behavior. Therefore, we refitted the form
factors according to this prescription. In the region $Q^2 \le 4 \GeV^2$,
however, the difference between our new fit and the one performed in
\cite{Lalakulich:2006sw} falls within the accuracy of the
experimentally extracted helicity amplitudes. To be on the safe side
for higher $Q^2$ values, further attempts to improve the fits of the
form factors (for example, in accordance to upcoming data on helicity
amplitudes) will be done within the framework of the arguments
presented in \cite{Vereshkov:2007xy}. The axial form factors are the
ones used in \cite{Lalakulich:2006yn} for the ``fast'' fall--off
case. Thus, we use the following form factors

\begin{equation}
\begin{array}{ll}
P_{33}(1232): &  C_3^{(p)}=\frac{2.14/D_V}{1+Q^2/4 M_V^2}, \quad 
		C_4^{(p)}=\frac{-1.56/D_V}{(1+Q^2/7.3 M_V^2)^2}, \quad 
		C_5^{(p)}=\frac{0.83/D_V}{(1+Q^2/0.95 M_V^2)^2},
\\[3mm]
	     &  C_i^{(n)}=C_i^{(p)}, \qquad C_i^{V}=C_i^{(p)},
\\[3mm]
             &   C_3^A=0, \quad C_4^A=-C_5^A/4, \quad  C_5^A=\frac{1.2/D_A}{1+Q^2/3M_A^2}, 
		\quad C_6^A=m_N^2 \frac{C_5^A}{m_\pi^2+Q^2}, 
\end{array}
\end{equation}
\begin{equation}
\begin{array}{ll}
P_{11}(1440): &   g_1^{(p)} = \frac{2.2/D_V}{1+Q^2/1.2 M_V^2}\left[ 1.+0.97\ln\left(1.+\frac{Q^2}{1\GeV^2}\right)\right], \quad
		  g_2^{(p)} = \frac{-0.76/D_V}{(1+Q^2/43 M_V^2)^2} \left[1 - 2.08 \ln\left(1+\frac{Q^2}{1\GeV^2}\right) \right],
\\[3mm]
		&  g_{i}^{(n)}=-g_{i}^{(p)}, \qquad g_{i}^{V}=g_{i}^{(n)}-g_{i}^{(p)},
\\[3mm]
		&  g_1^A=\frac{-0.51/D_A}{1+Q^2/3M_A^2}, \qquad   g_3^A=\frac{(M_R+m_N)m_N}{Q^2+m_\pi^2} g_1^A{}^{(P)},
\end{array}
\end{equation}
\begin{equation}
\begin{array}{ll}
D_{13}(1520):  &  C_3^{(p)}=\frac{2.95/D_V}{1+Q^2/8.0 M_V^2},      \quad 
                  C_4^{(p)}=\frac{-1.05/D_V}{(1+Q^2/17 M_V^2)^2},     \quad 
		  C_5^{(p)}=\frac{-0.48/D_V}{(1+Q^2/37 M_V^2)^2} .
\\[3mm]
		& C_3^{(n)}=\frac{-1.13/D_V}{1+Q^2/8.0 M_V^2},   \quad 
		  C_4^{(n)}=\frac{0.46/D_V}{(1+Q^2/17 M_V^2)^2},    \quad 
		  C_5^{(n)}=\frac{-0.17/D_V}{(1+Q^2/37 M_V^2)^2} ,
\\[3mm]
		&  C_{i}^{V}=C_{i}^{(n)}-C_{i}^{(p)},
\\[3mm]
		&  C_3^A=0, \quad C_4^A=0, \quad C_5^A=\frac{-2.1/D_A}{1+Q^2/3M_A^2}, \quad  C_6^A=m_N^2 \frac{C_5^A}{m_\pi^2+Q^2},
\end{array}
\end{equation}
\begin{equation}
\begin{array}{ll}
S_{11}(1535):   &  g_1^{(p)}=\frac{1.87/D_V}{1+Q^2/1.2 M_V^2} \left[1+7.07\ln\left(1+ \frac{Q^2}{1\GeV^2}\right)  \right] \ , \quad 
		   g_2^{(p)}=\frac{0.64/D_V}{(1+Q^2/17 M_V^2)^2} \left[1 + 1.0 \ln\left(1+\frac{Q^2}{1\GeV^2}\right)  \right] \ ,
\\[3mm]
		&  g_{i}^{(n)}=-g_{i}^{(p)}, \qquad g_{i}^{V}=g_{i}^{(n)}-g_{i}^{(p)},
\\[3mm]
		&  g_1^A=\frac{-0.21/D_A}{1+Q^2/3M_A^2}, \quad     g_3^A=\frac{(M_R-m_N)m_N}{Q^2+m_\pi^2} g_1^A \ .
\end{array}
\end{equation}

Here, $D_V=(1+Q^2/M_V^2)^2$ with $M_V=0.84\GeV$ and
$D_A=(1+Q^2/M_A^2)^2$ with $M_A=1.05\GeV$. The weak form factors
presented here are determined for the excitation of the $R^+$
resonance state, i.e. for neutrino scattering on a neutron. For the
excitation of the double charged states, which is possible for
isospin-3/2 resonances in neutrino--proton scattering, the isospin
relation gives an additional factor $\sqrt{3}$ for each form factor.

For the resonance widths we use the so called running widths $\Gamma_R(W)$, as they were presented in  Ref.~\cite{Lalakulich:2006sw}:
\begin{equation}
\Gamma_R(W)=\Gamma_R^0\left( \frac{p_\pi(W)}{p_\pi(M_R)} \right)^{2s_R},
\label{widths}
\end{equation}
where $s_R$ is the spin of the resonance, on--shell widths are $\Gamma_\Delta^0=0.12\GeV$,
$\Gamma_{P1440}^0=0.350\GeV$, $\Gamma_{D1520}^0=0.125\GeV$, $\Gamma_{S1535}^0=0.150\GeV$, and 
\[
 p_{\pi}(W)=\frac1{2W}\sqrt{(W^2-m_N^2-m_\pi^2)^2-4m_N^2 m_\pi^2} \ .
\]


\subsection{Definition of the  nuclear structure functions \label{defSF}}

For nuclear targets, the nuclear structure functions ${\cal W}_i^A$
can be defined in the standard manner by means of the expansion of the nuclear hadronic
tensor
\begin{equation}
 W^A_{\mu\nu}=
-g_{\mu\nu}{\cal W}_1^A 
+  \frac{p_\mu^A p_\nu^A}{M_A^2} {\cal W}_2^A 
- i\varepsilon_{\mu\nu\lambda\sigma} \frac{p_A^\lambda q^\sigma}{2 M_A^2} {\cal W}_3^A 
+  \frac{q_\mu q_\nu}{M_A^2} {\cal W}_4^A 
+  \frac{p_\mu^A q_\nu + p_\nu^A q_\mu}{M_A^2} {\cal W}_5^A \; ,
\label{Wmunu-nucleus}
\end{equation}
where $p^{A, \, \mu}=(M_A, \vec 0)$ is the four--momentum of the
target nucleus with mass $M_A$ in the laboratory frame of reference.

In the impulse approximation we are dealing with the bound nucleon as
a target, so we must relate the one--bound--nucleon structure
functions introduced in the previous section to nuclear ones.  We
follow the prescription of Ref.~\cite{Atwood:1972zp} and express the
nuclear structure functions in terms of the nucleon ones in terms of a
convolution of the type
\begin{equation}
 W^A_{\mu\nu}= \sum\limits_{\alpha} \int d^3 p \; (2j_\alpha+1) n_\alpha(p)  (W_{\mu\nu (\alpha)}^p + W_{\mu\nu (\alpha)}^n)  \ ,
\label{WAW}
\end{equation}
where $\alpha$ extends over single--particle shells in the target
nucleus and $2j_\alpha+1$ specifies their occupancies. 

It is worth stressing that in the original paper \cite{Atwood:1972zp}
as well as in \cite{Ferree:1995fb} an additional phase--space
correction factor $E_p/m_N$ is introduced in the
expression~(\ref{WAW}) to preserve the space volume under Lorentz
transformation. Since we construct a momentum distribution from wave
functions normalized as $u_\alpha ^\dagger u_\alpha = 1$ for each
shell $\alpha$, our correction factor must be equal to~$1$.

Substituting (\ref{Wmunu-nucleon}) and (\ref{Wmunu-nucleus}) in (\ref{WAW}), one arrives at
\begin{equation}
\begin{array}{l} \displaystyle
{\cal W}_1^A(Q^2,\nu)=\sum\limits_{\alpha} {\cal W}_1^{(\alpha)}(Q^2,\nu)
=\sum\limits_{\alpha} \int d^3 p \; (2j_\alpha+1) n_\alpha(p) 
\biggl[ {\cal W}_1(Q^2,\nu, p^2) 
+ {\cal W}_2(Q^2,\nu, p^2)\frac{|\vec p|^2 -p_z^2}{m_N^2} \biggr] 
\ , 
\\[4mm] \displaystyle
{\cal W}_2^A(Q^2,\nu) =\sum\limits_{\alpha} {\cal W}_2^{(\alpha)}(Q^2,\nu)
=\sum\limits_{\alpha} \int d^3 p \; (2j_\alpha+1) n_\alpha(p) 
{\cal W}_2(Q^2,\nu, p^2) 
 \left[\frac{|\vec p|^2 -p_z^2}{m_N^2}\frac{Q^2}{q_z^2} 
      + \left( \frac{(p\cdot q)}{m_N\nu} \right)^2 \left( 1+ \frac{p_z}{q_z}\frac{Q^2}{(p\cdot q)} \right)^2 
\right]. 
\end{array}
\label{calWA}
\end{equation}
This prescription guarantees, that as $Q^2$ tends to zero, the
longitudinal structure function ${\cal W}_L$ also tends to zero as
expected for the real photon:
\begin{equation}
 \lim\limits_{Q^2\to 0} \left[ \frac{\nu^2}{Q^2} {\cal W}_2^A (Q^2,
 \nu) - {\cal W}_1^A (Q^2, \nu) \right] =0 \; .
\end{equation}

In neutrino experiments one can also measure the ${\cal W}_3$
structure function, for which our definition gives:
\begin{equation}
{\cal W}_3^A(Q^2,\nu)=\sum\limits_{\alpha} \int d^3 p \; (2j_\alpha+1)
n_\alpha(p) {\cal W}_3(Q^2,\nu, p^2) \frac{M_A}{m_N^2} \frac{p^0 q^z -
\nu p^z }{q_z} \; .
\end{equation}
Note that ${\cal W}_3$ depends on the nucleus mass $M_A$. Realizing
that the Bjorken variable for a nucleus ($x_A=Q^2/2 M_A \nu$) differs
from the one for a nucleon ($x=Q^2/2m_N\nu$), the function that is
independent of $M_A$ is $x_A F_3^A$:
\begin{equation}
x_A F_3^A = 
  \sum\limits_{\alpha} \int d^3 p \; (2j_\alpha+1) n_\alpha(p)
 xF_3(Q^2,\nu, p) \frac{1}{m_N} \frac{p^0 q^z - \nu p^z }{q_z} \; .
\end{equation}
According to the definition (\ref{calWA}), $F_2^A=\nu {\cal W}_2^A$ and
$x_A F_1^A = x_A M_A {\cal W}_1^A$ are also independent on $M_A$.

Within the adopted approach there is no unambiguous recipe
for deciding whether one should keep $m_N$ in the denominators
of~(\ref{calWA}) or replace it with some effective mass, that corrects
for the binding energy. For the numerical calculations presented here,
we have opted to use the expression ~(\ref{calWA}) and interpret the
$m_N$ as the free nucleon mass.

The integration over $d^3p=|\vec{p}|^2 \, d|\vec{p}| \; d\cos\gamma_p
\; d\varphi_p$ in Eq.~(\ref{calWA}) is performed in the following
way. Integration over the azimuthal angle $d\varphi_p$ gives $2\pi$,
since no structure function depends on it. The phase space in the
plane determined by the absolute momentum value $|\vec{p}|$ and polar
angle $ \gamma _p$ is restricted by the condition $W^2>W_{min}^2$. For one--pion
production one has that $W_{min}=m_N+m_\pi$. For a a bound nucleon 
this condition translates into
\begin{equation}
 p_0^2-|\vec{p}|^2+2p_0\nu
-2|\vec{p}|\sqrt{Q^2+\nu^2} \cos\gamma_p - Q^2 > W_{min}^2 \; .
\label{phasespace}
\end{equation}
When performing the $d^3 {p}$ integrations, the above condition   
determines the boundaries of the absolute bound-nucleon momentum 
$| \vec{p} | $ for a given $\cos \gamma _p$, $Q^2$ and $\nu$: 
\begin{equation}
|\vec{p}|_{\pm} = -\sqrt{Q^2+\nu^2}\cos \gamma_p 
\pm  \sqrt{(Q^2+\nu^2)\cos^2\gamma_p +p_0^2 +2 p_0\nu -Q^2 -W_{min}^2 } \ .
\label{p_pm}
\end{equation}

The sign of the quantity $W_{min}^2 + Q^2 - p_0^2 - 2p_0\nu$
discriminates between two classes of kinematic conditions.  In what
follows we provide a discussion of the values of $p_{min}$ and
$p_{max}$ in the phase-space integration $ \int _{p_{min}} ^{p_{max}}
dp $ for a positive and negative sign of $W_{min}^2 + Q^2 - p_0^2 -
2p_0\nu$. For
\begin{equation}
 W_{min}^2 + Q^2 - p_0^2 - 2p_0\nu<0 \ ,
\label{also-free}
\end{equation}
the $|\vec{p}|_{-}$ calculated according to ({\ref{p_pm}) is negative, so one
should take $p_{min}(Q^2,\nu,\cos\gamma_p)=0$. This means that the
phase space (\ref{also-free}) is accessible for a nucleon with
arbitrarily small three-momentum, including $|\vec{p}|=0$, as is the
case for a free nucleon. When the condition (\ref{also-free}) is
fullfilled, $p_{max}(Q^2,\nu,\cos\gamma_p) =
|\vec{p}|_{+}(Q^2,\nu,\cos\gamma_p)$ for all polar angles $\gamma_p$.

Increasing the phase space for the bound nucleon does not necessarily
imply that the cross section grows, because each point in the phase
space gets weighted with a momentum distribution of the type shown in
Fig.~\ref{fig:n_alpha}. Cross sections and structure functions for
high $|\vec{p}|$ are strongly suppressed and the major contributions
stem from the momenta inside the Fermi sphere.

For 
\begin{equation}
 W_{min}^2 + Q^2 - p_0^2 - 2p_0\nu>0 \ ,
\label{bound-only}
\end{equation}
the $|\vec{p}|_{\pm}$ are only defined for backward moving target
nucleons. The restrictions on $\cos\gamma_p$ for given $Q^2$ and $\nu$
come from the condition
\begin{equation}
(Q^2+\nu^2)\cos^2\gamma_p +p_0^2 +2 p_0\nu -Q^2 -W_{min}^2 > 0 \ ,
\end{equation}
which gives
\begin{equation}
-1 < \cos\gamma_p(Q^2,\nu)  < -\sqrt{\frac{W_{min}^2 + Q^2 - p_0^2 - 2p_0\nu}{Q^2+\nu^2}} \ .
\label{bound-cosgamma}
\end{equation}
Since the minimal value of the three-momentum $|\vec{p}|_{-}$ is
positive in this case, the accessibility to this $(Q^2,\, \nu)$ region
crucially depends on a nucleon already moving, which is only possible
for a bound nucleon.  This region of phase space grows in importance
with increasing $Q^2$.

For $Q^2=0.1\GeV^2$ and different $\nu$, the typical phase spaces
available for a $1s^{1/2}$ proton in carbon are shown in
Fig.~\ref{fig:phasespace}.  We use polar coordinates for the variables
$|\vec{p}| $ and $ \gamma_p$.  The left (right) panel corresponds with
the condition (\ref{bound-only}) ((\ref{also-free})). For each $Q^2$
and $\nu$, thick lines represent $|\vec{p}|_{+}$ and thin lines
$|\vec{p}|_{-}$. The points where the $|\vec{p}|_+$ and $|\vec{p}|_- $
lines coincide correspond to the upper boundary on $\cos\gamma_p$, as
calculated in Eq.(\ref{bound-cosgamma}). Remark that the available
phase space in ($|\vec{p}|, \cos \gamma_p)$ is contained within a circle. At
$\nu=0.4\GeV$ the left part of the circle is not shown because it
corresponds to values of $|\vec{p}|$ larger than 1 GeV. The momentum
distribution of nucleons in nuclei will reduce those contributions to
negligible proportions.

\begin{figure}[hbt]
\epsfig{figure=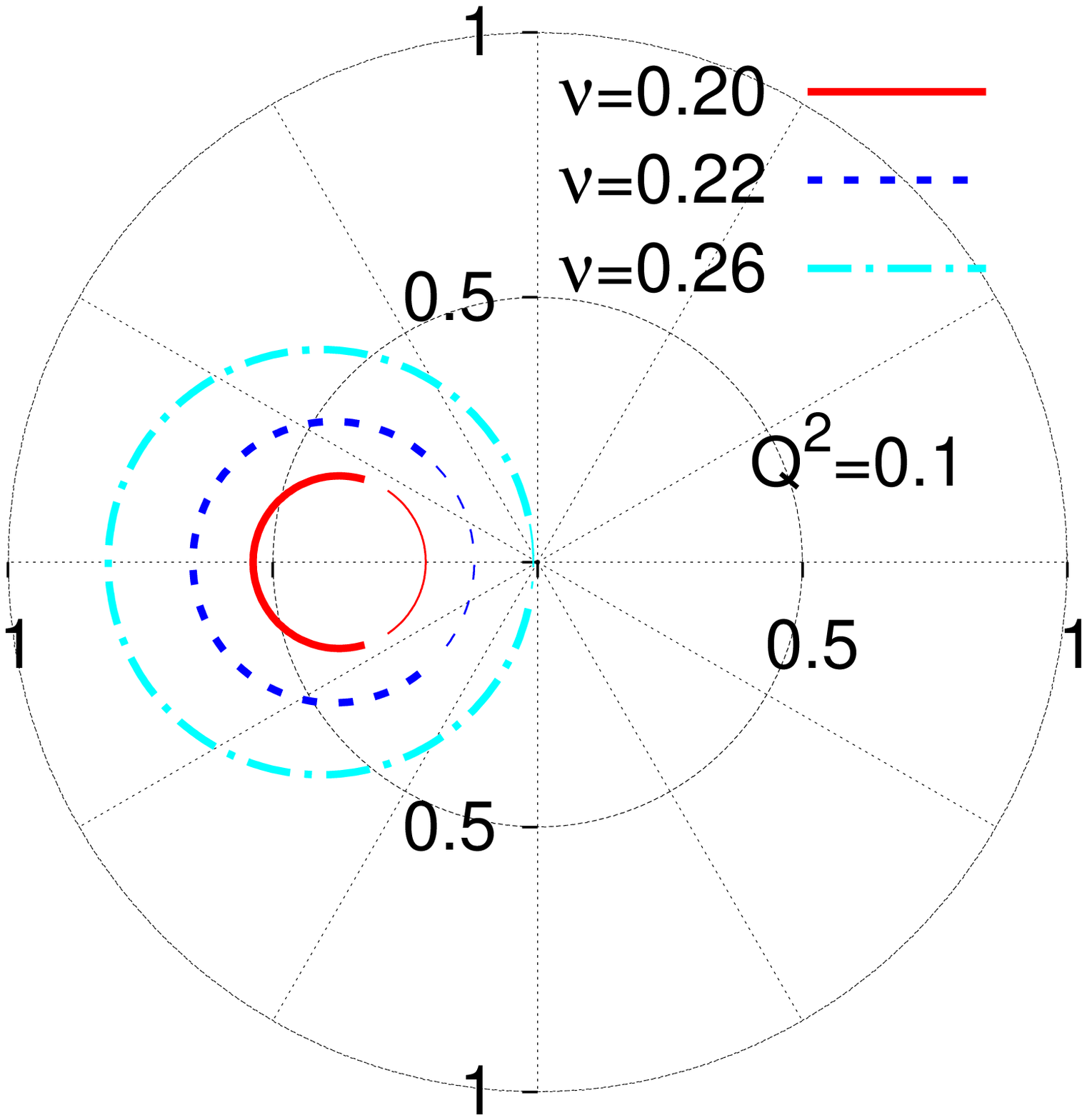,angle=-90,width=0.49\textwidth}
\hfill
\epsfig{figure=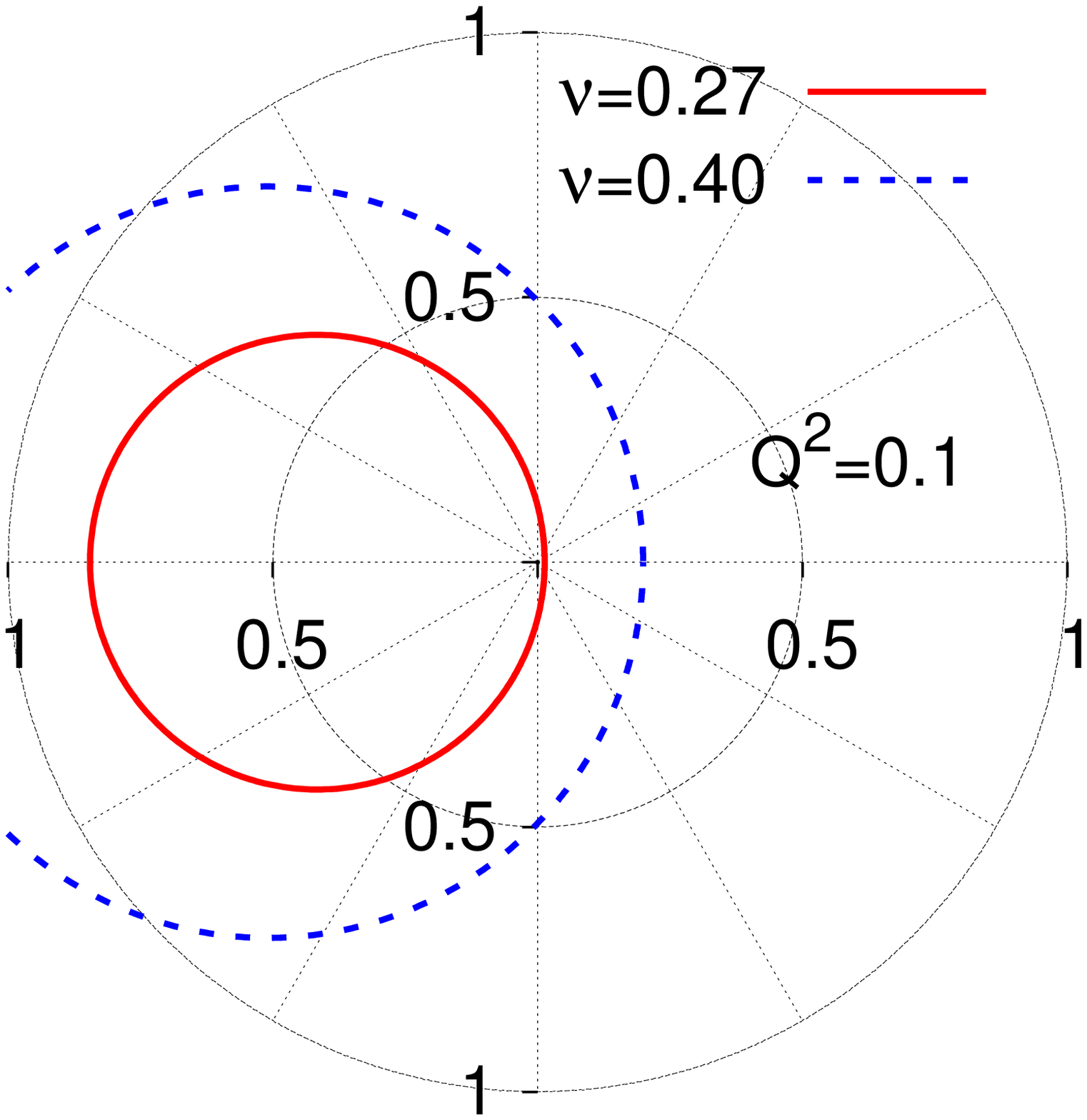,angle=-90,width=0.49\textwidth}
\caption{Sketch of the available $(|\vec{p}| \,  ,\gamma_p)$  phase space
  in  polar coordinates. We consider a $1s^{1/2}$ proton in carbon for
  $Q^2=0.1 \GeV^2$ and different $\nu$. The $ |\vec{p} | $ is
  expressed in GeV. The left (right) panel corresponds with kinematics
  conditions obeying the condition of Eq.~(\ref{bound-only}) (of Eq.~(\ref{also-free})).}
\label{fig:phasespace}
\end{figure}

The phase space collapses to one point $\cos\gamma_p=-1$,
$|\vec{p|}=\sqrt{Q^2+\nu_{min}^2}$ for
$\nu_{min}=-p_0+W_{min}$. Remark that for a bound nucleon the minimal
value of $\nu$ does not depend on $Q^2$. Physically this means that for any $Q^2$
there is a bound nucleon moving backward fast enough to fulfill the
requirement $(p+q)^2>W_{min}^2$. Thus, contrary to the free nucleon
case, for the off-shell nucleon the pion production threshold is
defined in terms of $\nu$ rather than invariant mass and, strictly
speaking, is independent of $Q^2$.

At high $Q^2$, however, using $\nu_{min}=-p_0+W_{min}$ is not
convenient for calculations, because all observables are strongly
suppressed for large $\vec{p}$ inspite of the fact that the phase
space is available. In our numerical calculations we have not
considered nucleon momenta beyond three times the Fermi momentum. 
%
%
We stress that the phase space boundaries derived here depend
on our assumption about the form of the four--momentum for the bound
nucleon, which was taken as $p^\mu=(m_N-e_{\alpha}, \vec{p})$.

%

\subsection{DIS region and scaling variable \label{DIS}}

In the kinematical regime of high $Q^2$ and $\nu$, the so-called 
Bjorken limit, the
structure functions depend only on the Bjorken variable
$x=Q^2/2m_N\nu$ when one neglects higher-twist effects. This phenomenon of no observed $Q^2$ dependence for a
fixed $x$ value is called Bjorken scaling. At these energies, the
lepton scattering on nucleons and nuclei is dominated by deep
inelastic scattering with a multiple--particle hadronic final state. 
Deep inelastic scattering on nuclei was intensively studied
experimentally since the sixties. This experimental information will be
used as DIS input for our investigation.

For electron--carbon scattering, $F_2$ was measured by the BCDMS
Collaboration~\cite{Bollini:1981cr,Benvenuti:1987zj} for
$30\GeV^2<Q^2<200\GeV^2$. We choose several sets of data at different
$Q^2=Q^2_{DIS}$: $30, \, 45$ and $50 \GeV^2$. As expected from Bjorken
scaling, for most of the $x$ region the data coincide with an accuracy
better than $5\%$. For iron, the neutrino scattering results are
available from the CCFR~\cite{Seligman:1997mc} and
NuTeV~\cite{Tzanov:2005kr} collaborations.

Scaling structure functions are conventionally plotted against the
Bjorken variable $x$. Violation of Bjorken scaling comes from
target-mass corrections and higher-twist effects. In the scaling
region, the Nachtmann variable $\xi= 2x/(1+\sqrt{1 + 4 m_N^2
  x^2/Q^2})$ was shown \cite{Nachtmann:1973mr,Georgi:1976ve} to be a
better alternative, because it implicitly includes the kinematical
part of the target-mass correction, which can be important at large
$x$ and low $Q^2$. Expanding the inverse of this variable in a power
series of $1/Q^2$, we recover the variable $1/\xi\approx \omega'=(2m_N
\nu + m_N^2)/Q^2$, used by Bloom and Gilman in their pioneering work
on duality. For large $\nu$, one has $\omega'\approx 1/x$.

%
%
%


\section{Duality in electroproduction \label{electrons}}

In the case of an isoscalar target nucleon, and for $Q^2>0.5\GeV^2$,
it was shown \cite{Lalakulich:2006yn} that Bloom--Gilman duality holds 
at the level of $20\%$. 
Here, we compute the nuclear structure functions $F_2^A$  and $x_AF_1^A$
along the model outlined in Section II.  The results of our 
description of the resonance region in terms of hadronic degrees of
freedom are then compared to DIS data. 

%
%

In Figure~\ref{fig:C12-em}, the bound--nucleon structure functions
$F_2$ and $2x F_1$ for a proton in the $1s^{1/2}$ and $1p^{3/2}$
carbon shells are contrasted with the structure functions for a free
proton. They are plotted versus the Nachtmann variable $\xi$ for
$Q^2=0.2, \, 0.85$, and $2.4 \GeV^2$, with the largest $Q^2$ curves
covering the largest $\xi$ values. Similar to the free--nucleon case,
for a fixed $Q^2$, the peak at larger $\xi$ corresponds to the
$\Delta$ resonance and the peak at smaller $\xi$ corresponds to the
second resonance region. One can easily notice the effect of smearing:
the two resonance regions are distinguishable only at low $Q^2$. Fermi
smearing proceeds differently for different shells, which in turn
introduces an additional averaging when summing over shells.
One can also observe that the bound--nucleon curves extend to higher
$\xi$ values than the free nucleon ones. This additional contribution
comes from the phase space at low $\nu$ values, which is shown in the
left panel of Fig.~\ref{fig:phasespace} and discussed in
Section~\ref{defSF}. At high $\xi$ the $F_2$ and $2 x F_1$ for the $1s^{1/2}$
shell are significantly lower than for the $1p^{3/2}$ shell. At high $\xi$ 
the phase space extends to relatively large bound-nucleon momenta.
For those momenta the momentum
distribution for a shell close to the Fermi surface (like $1p^{3/2}$
in carbon) is larger than for a deep-lying shell (like $1s^{1/2}$ in
carbon).  

%


%
%
%

\begin{figure}[htb]
\begin{minipage}[c]{0.49\textwidth}
        \epsfig{figure=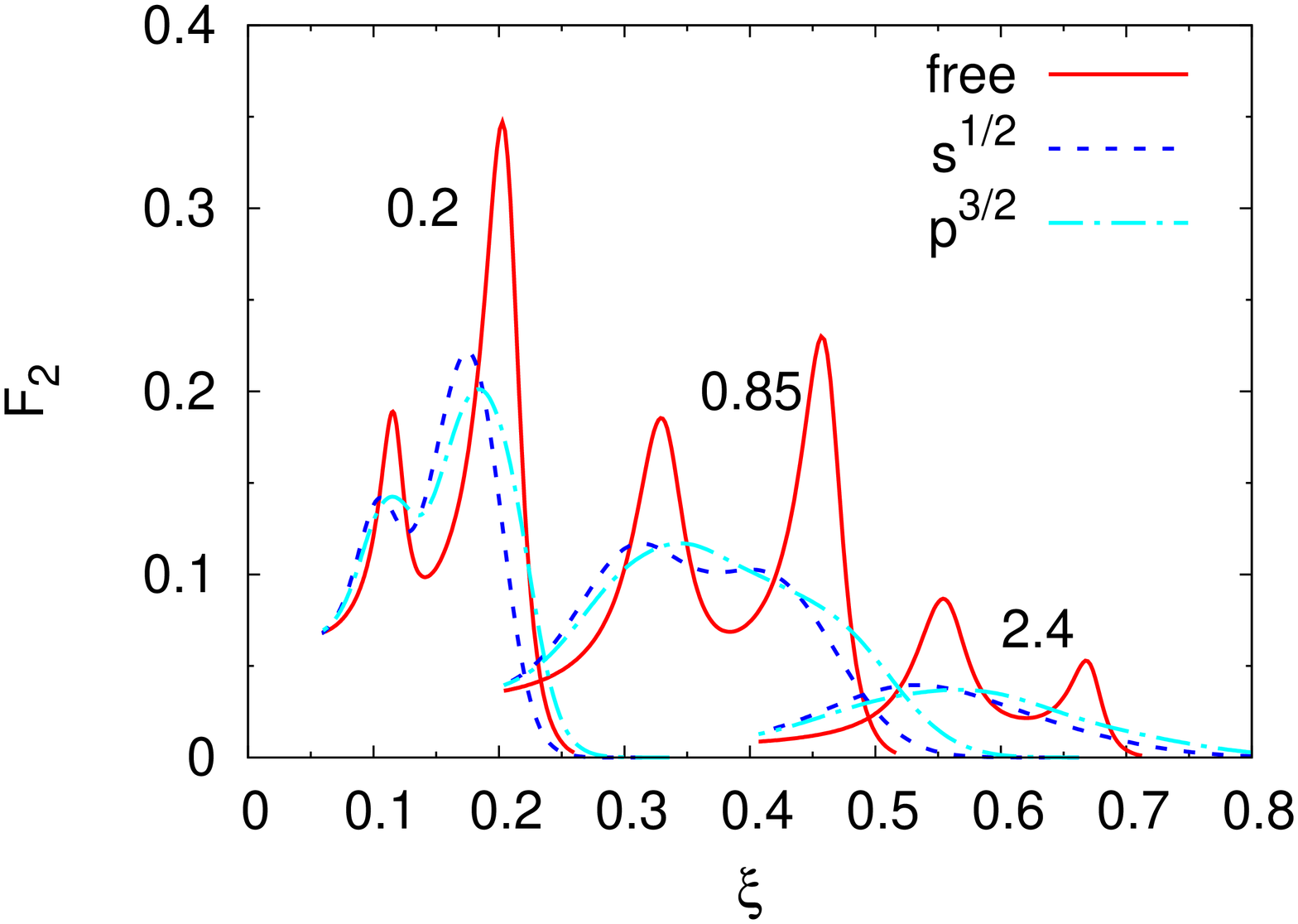,angle=-90,width=\textwidth}
\end{minipage}
\hfill
\begin{minipage}[c]{0.49\textwidth}
        \epsfig{figure=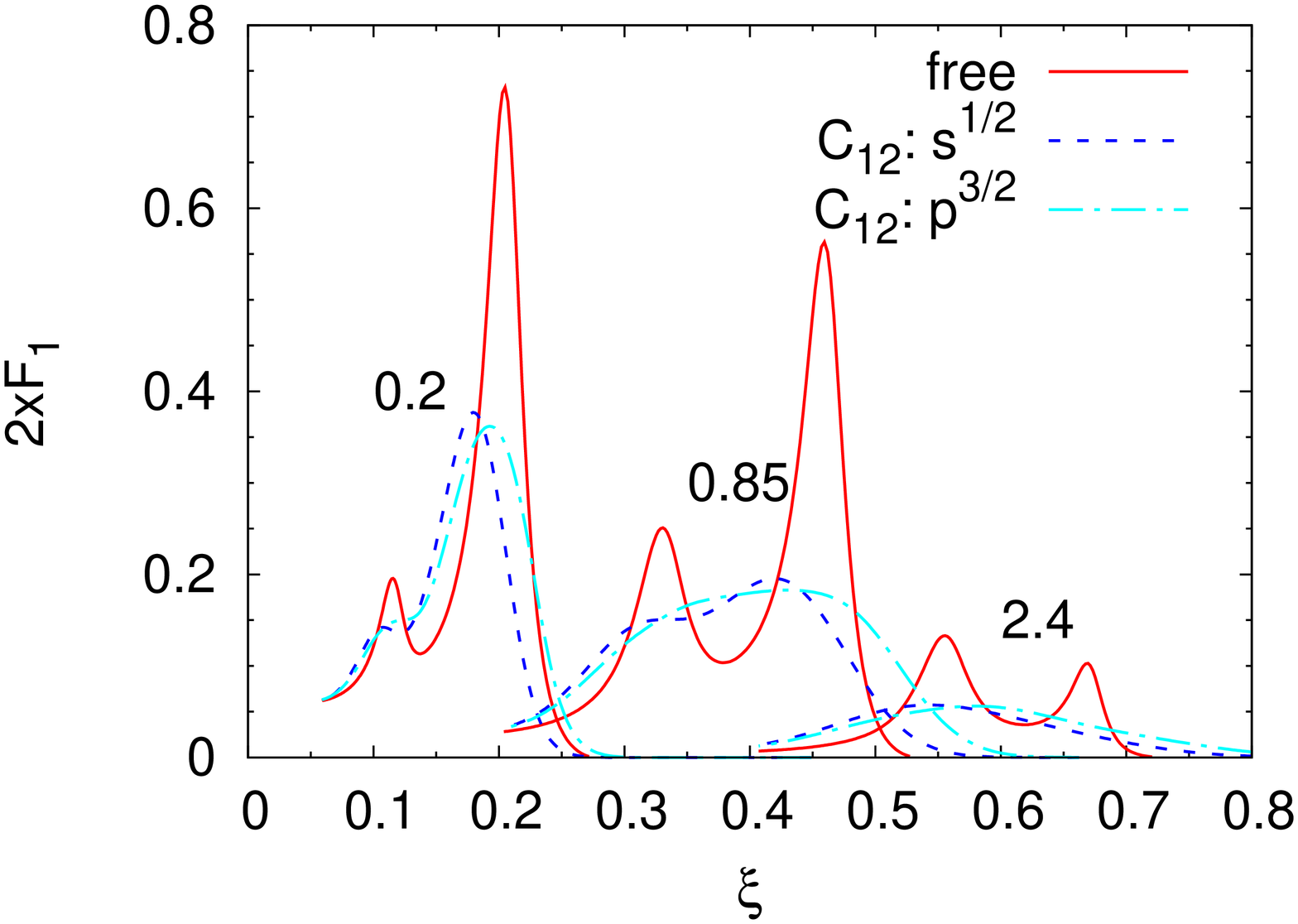,angle=-90,width=\textwidth}
\end{minipage}
        \caption{(color online) Structure functions $F_2$ (left) and $2xF_1$ (right) for a free proton (solid curve), for $1s^{1/2}$ (dashed curve) and $1p^{3/2}$ (dash--dotted curve) protons in ${}^{12}$C. The three sets of curves correspond to $Q^2=0.2$, $0.85$, and $2.4\GeV^2$} 
\label{fig:C12-em}
\end{figure}

Fig.~\ref{fig:F2-C12-em} shows the carbon structure function per
nucleon $F_2^{e{}^{12} C}/A$ in the resonance region for several $Q^2$
values, from $0.45$ to $3.3 \; {\mathrm{GeV}}^2$. When investigating
duality for a free nucleon, we took the average over free proton and
neutron targets, thus considering the isoscalar structure
function. Since the carbon nucleus contains an equal number of protons
and neutrons, averaging over isospin is performed automatically. At
$Q^2=0.45 \GeV^2$, the $\Delta$ peak is pronounced and can still be
distinguished from the second resonance peak, which is also
visible. At higher $Q^2$ one cannot distinguish the resonance
structure anymore and the first and second resonance region merge into
one broad peak. 
\begin{widetext}
\begin{figure}[h!bt]
\begin{minipage}[c]{0.49\textwidth}
        \epsfig{figure=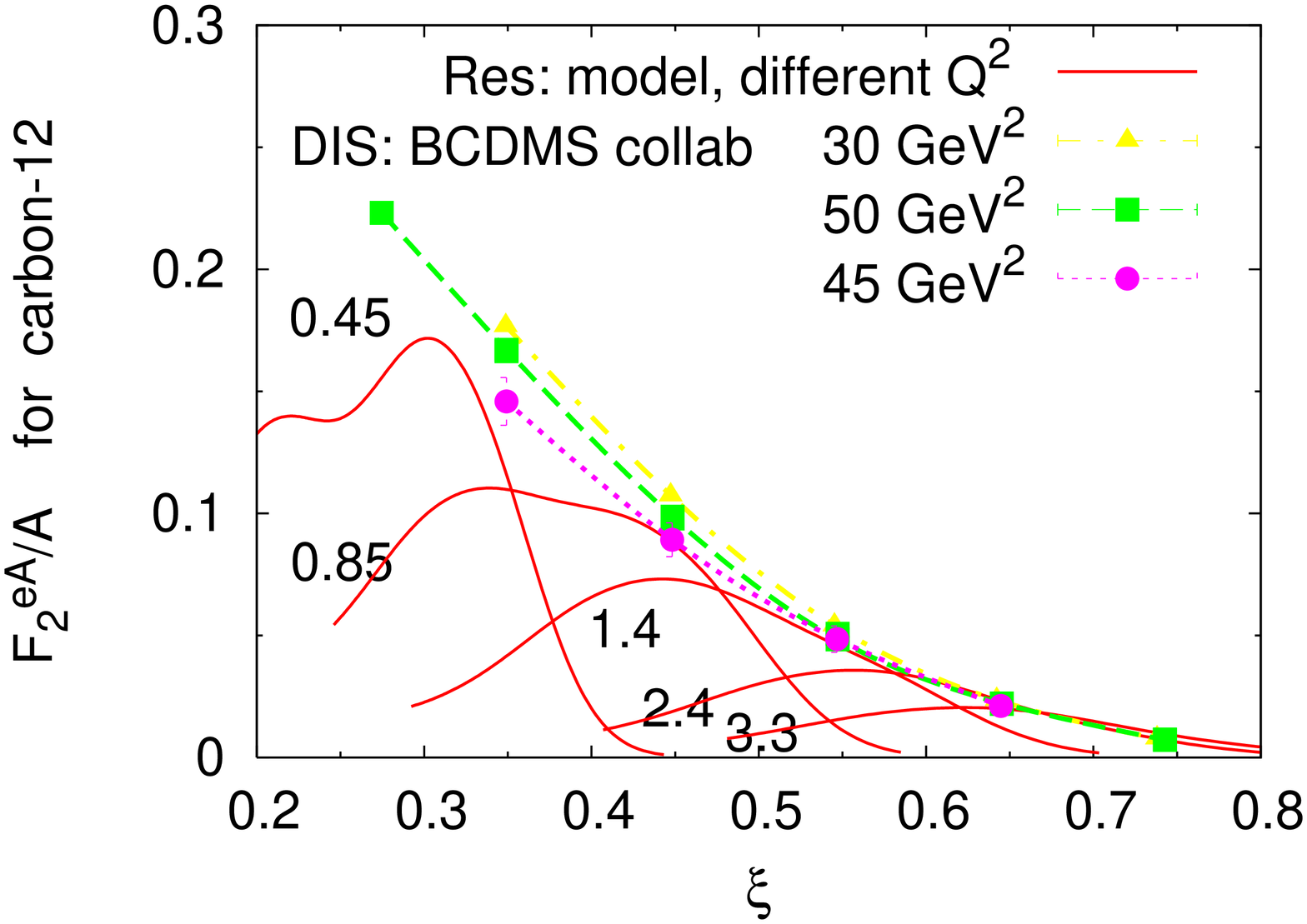,angle=-90,width=\textwidth}
\end{minipage}
\hfill
\begin{minipage}[c]{0.49\textwidth}
        \epsfig{figure=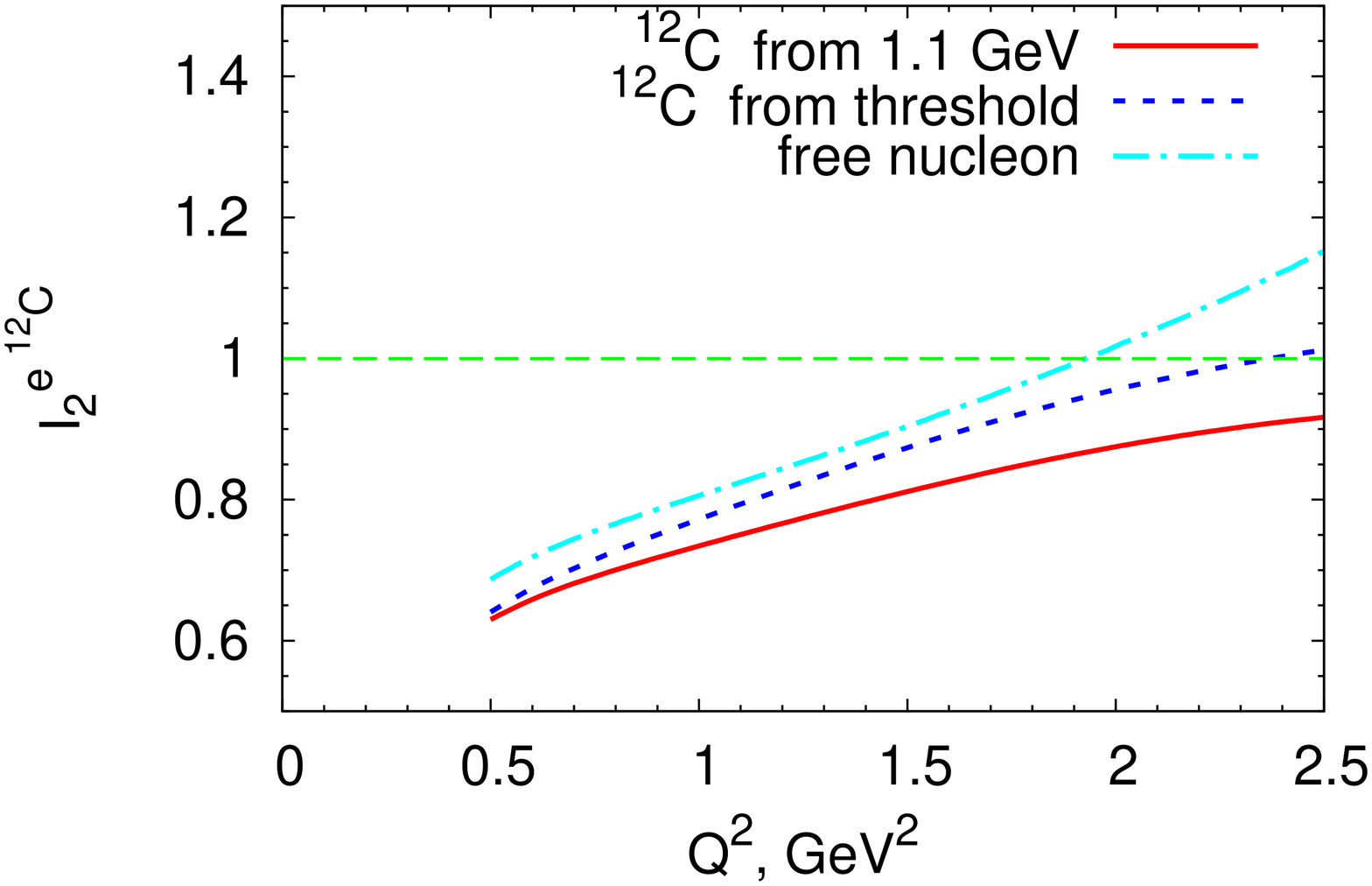,angle=-90,width=\textwidth}
\end{minipage}
        \caption{(color online) Duality for the $F_2^{e {}^{12}C}$
 structure function.  (Left) Resonance curves $F_2^{e {}^{12}C}/12$ as
 a function of $\xi$, for $Q^2 = 0.45, 0.85, 1.4, 2.4$ and $3.3\;
 {\mathrm{GeV}}^2$ (indicated on the spectra), compared with the
 experimental data \cite{Bollini:1981cr,Benvenuti:1987zj} in the DIS
 region at $Q^2_{DIS}=30$, $45$ and $50\GeV^2$.  (Right) Ratio $I_2$
 defined in Eq.(\ref{eq:Int}) for the free nucleon (dash-dotted line),
 and $^{12}$c.  We consider the under limits determined by
 $\tilde{W}=1.1 \GeV$ (solid line) and by the threshold value (dotted
 line).}
\label{fig:F2-C12-em}
\end{figure}
\end{widetext}
In the left panel of Fig.~\ref{fig:F2-C12-em}, the resonance structure
functions are compared with data obtained by the BCDMS
Collaboration~\cite{Bollini:1981cr,Benvenuti:1987zj} in muon--carbon
scattering in the DIS region ($Q^2\sim 30-50 \;
{\mathrm{GeV}}^2$). They are shown as experimental points connected by
smooth curves. For different $Q^2$ values, the curves agree within
$5\%$ in most of the $\xi$ region, as expected from Bjorken
scaling. One observes that, as $Q^2$ increases, the resonance peaks
decrease in height and slide along the DIS curve. This means that
global duality holds for electron scattering on nuclei.  To
characterize local duality, we consider the ratio of the integrals of
the resonance (res) and DIS structure functions
\begin{equation}
I_i(Q^2) =
\frac{ \int_{\xi_{\rm min}}^{\xi_{\rm max}} d\xi\
    {\cal F}_i^{(\rm res)}(\xi,Q^2) }
     { \int_{\xi_{\rm min}}^{\xi_{\rm max}} d\xi\
    {\cal F}_i^{(\rm DIS)}(\xi,Q^2_{DIS}) }\ ,
\label{eq:Int}
\end{equation}
where ${\cal F}_i$ denotes $F_2^A$ or $x_A F_3^A$ (used later for neutrino scattering). The value $Q^2_{DIS}$ is
taken as the actual $Q^2$ value for a given experimental data set. For
electron--carbon scattering we choose the data set
\cite{Benvenuti:1987zj} at $Q^2_{DIS}=50 \GeV^2$, because it covers
most of the $\xi$ region.
For a proton target \cite{Niculescu:2000tk}, the integration limits for
$\xi$ are conventionally chosen equal for both integrals and are
defined in such a way as to cover the first and second resonance
regions for each $Q^2$. For a free nucleon, this requirement is
written as \cite{Lalakulich:2006yn}
\begin{equation}
\xi_{\rm min}^{N} = \xi(W=1.6~{\rm GeV},\, Q^2), \qquad  
\xi_{\rm max}^{N} = \xi(W=1.1~{\rm GeV},\, Q^2), 
\label{11-16}
\end{equation}
where the invariant mass for a free nucleon can be expressed in terms
of $\nu$ and $Q^2$ as $W^2=(p+q)^2=m_N^2 +2m_N\nu -Q^2$. The upper
value $W=1.6 \GeV$ is chosen in such a way as to cover the mass range
of the four resonances taken into account, the heaviest one with the
mass $M_R=1.535 \GeV$. The lower value $W=1.1 \GeV$ is chosen close to
the pion--production threshold $W_{thr}=1.08\GeV$.
In a nuclear target, the invariant mass of the struck nucleon 
depends on the initial momentum of the target nucleon. On the other
hand, the structure functions, as well as other observables, are
defined as integrals over the initial nucleon momentum. This prevents
one from using $W$ in defining the integration limits. One needs an
alternative variable, which can be easily determined from the lepton
kinematics. 

Experimentally one often (see, for example, \cite{Sealock:1989nx})
``defines'' the effective variable $\tilde{W}$ by the relation
$\tilde{W}^2=m_N^2+2m_N\nu-Q^2$. Notice that $\tilde{W}$ is only an 
invariant for $\vec{p}=0$. However, it gives a reasonable
feeling of the invariant mass region involved in the problem. In
particular, the resonance curves presented in all figures are plotted in the
region from the pion--production threshold up to $\tilde{W}=2\GeV$.
As was illustrated in Fig.~\ref{fig:phasespace}, bound backward-moving
nucleons allow  lower $\nu$ values beyond the free--nucleon limits. 
Thus, as discussed at the end of Secton~\ref{defSF}, the
threshold for the structure functions is now defined in terms of $\nu$
or $\tilde{W}$,  rather than $W$.  
Hence, we consider two different cases in choosing the
$\xi$ integration limits for the ratio (\ref{eq:Int}). First, for a
given $Q^2$, we choose the $\xi$ limits as in
Eq.~(\ref{11-16}). That amounts to defining them by the condition
\begin{equation}
\xi_{\rm min} = \xi(\tilde{W}=1.6~{\rm GeV},\, Q^2), \qquad  
\xi_{\rm max} = \xi(\tilde{W}=1.1~{\rm GeV},\, Q^2) \ .
\label{11-16-}
\end{equation}
We refer to this choice as integrating ``from 1.1 GeV''. The integration
limits for the DIS curve always correspond to this choice.  As a
second choice, for
each $Q^2$ we integrate the resonance curve from the threshold, that
is from as low $\tilde{W}$ as achievable for the nucleus under
consideration. This corresponds to the threshold value at higher $\xi$
and is referred to as integrating ``from threshold''. With this choice we 
guarantee that the extended kinematical regions typical for resonance 
production from nuclei are taken into account.
Since there is no natural threshold for the $\xi_{min}$, for
both choices it is estimated from $\tilde{W}=1.6\GeV$, as defined in
Eq.~(\ref{11-16-}).

%
%

The results for the ratio in Eq.~(\ref{eq:Int}) are shown in the right
panel of Fig.~\ref{fig:F2-C12-em}. The curve for the isoscalar
free-nucleon case is the same as in Ref.~\cite{Lalakulich:2006yn} with
the ``GRV'' parameterization for the DIS structure function. One can
see that the carbon curve obtained by integrating ``from threshold''
lies above the one obtained by integrating ``from 1.1 GeV'', the
difference increasing with $Q^2$. This indicates that the threshold
region becomes more and more significant, as one can see from
Fig.~\ref{fig:C12-em}.  
The closer the ratio (\ref{eq:Int}) gets to 1, the higher the accuracy
of local duality is. Our calculations for a carbon target show that:
1) the ratio grows with $Q^2$, just like in the isoscalar free-nucleon
case; 2) the ratio is lower than the free-nucleon value for both
choices of the integration limits. This means that the integrated
resonance contribution is always smaller than the integrated DIS one. 
\begin{figure}[h!b]
        \epsfig{figure=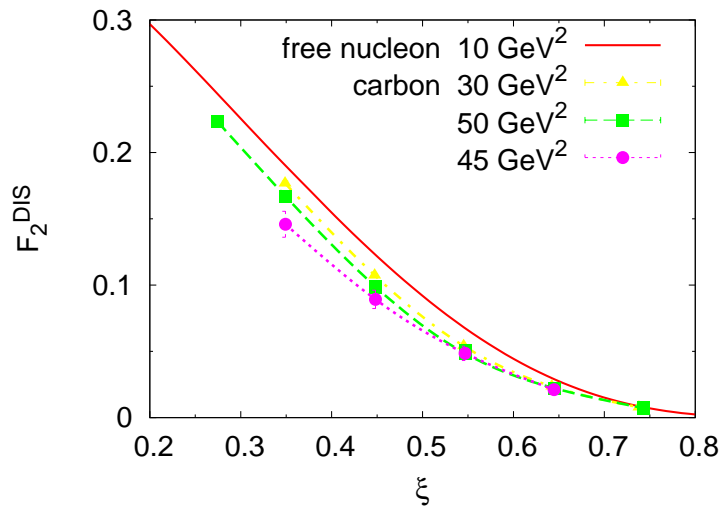,angle=-90,width=0.55\textwidth}
        \caption{Electromagnetic structure functions $F_2$ in DIS
          region for a free isoscalar nucleon as obtained via the GRV
          parameterization at $Q^2=10\GeV^2$ (solid curve) and for a
          carbon nucleus as measured experimentally at
          $Q^2=30, \, 45$ and $50\GeV^2$.}
\label{fig:F2-C-em-DIS}
\end{figure}
In search for an explanation for this discrepancy, we compare how
nuclear effects influence the resonance and DIS curves. As it was
illlustrated in Fig.~\ref{fig:C12-em}, the nuclear effects suppress
the resonance peaks by 40\%-50\%, broaden them and shift them to lower
$\xi$ values. The experimental DIS values for the carbon nucleus, on
the other hand, are only $5\%-10\%$ lower than the DIS curve for the
free isoscalar nucleon.  This is illustrated in
Fig.~\ref{fig:F2-C-em-DIS}, where the DIS structure function $F_2$ for
a carbon target is compared to the GRV parameterization for the free
isoscalar nucleon at $Q^2=10\GeV^2$. In conclusion, one can say that
nuclear effects have a much more dramatic effect in the resonance
region than in the DIS regime.


Similar calculations can be done for other nuclei. First of all, it
would be interesting to compare an isoscalar nucleus with a nucleus
with neutron excess. We show the structure functions $(F_2^{e\,
{}^{56}\! Fe}/56)$ and $(F_2^{e\, {}^{52}\! Fe}/52)$ versus
$\tilde{W}$ for several $Q^2$ values in the left panel of
Fig.~\ref{fig:F2-FeC-em}. The structure functions for $^{52}$Fe are
only marginally higher than those for $^{56}$Fe. This can be explained
by the fact that the electromagnetic $\Delta$-production cross section
is equal for proton and neutron targets. In the second resonance
region, the cross sections on the proton are typically $5\%-30\%$
higher than those for the neutron. The overall effect, however, is
hardly visible for an excess of 4 neutrons out of 56 nucleons.
\begin{figure}[h!b]
\begin{minipage}[c]{0.49\textwidth}
        \epsfig{figure=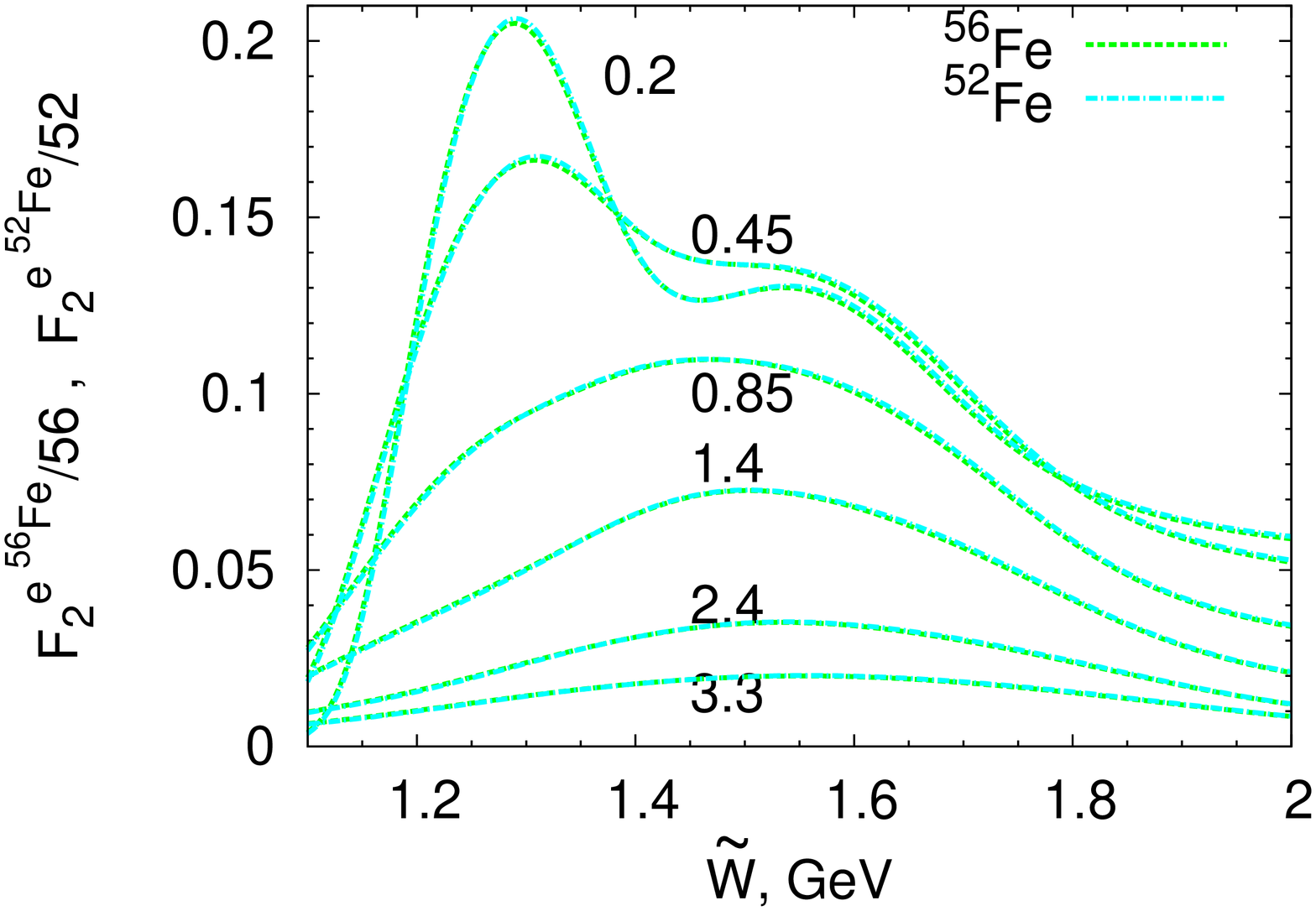,angle=-90,width=\textwidth}
\end{minipage}
\hfill
\begin{minipage}[c]{0.49\textwidth}
        \epsfig{figure=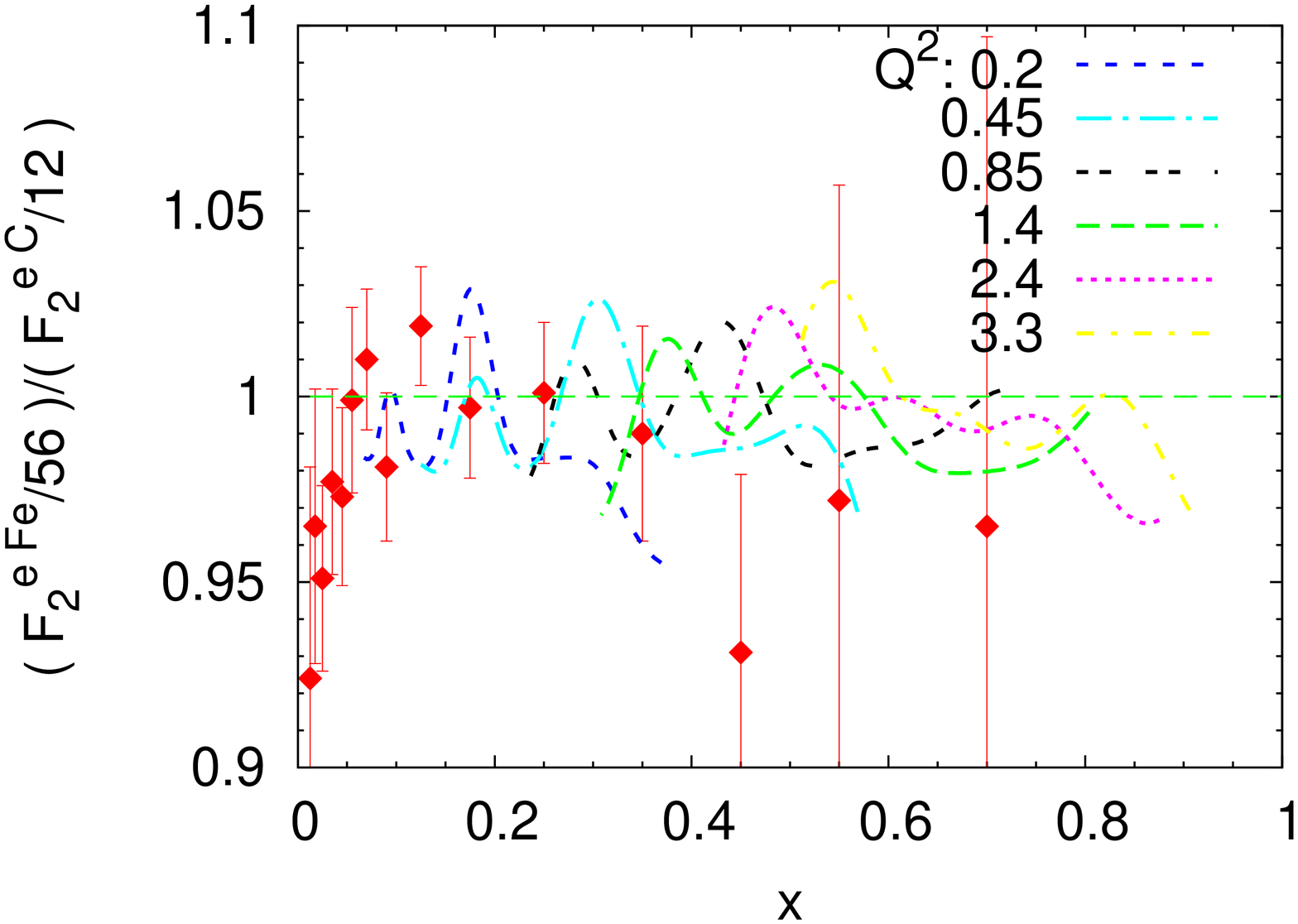,angle=-90,width=\textwidth}
\end{minipage}
        \caption{(color online) 
        (Left) Structure functions $F_2$ for electroproduction on iron-52 and iron-56 versus $\tilde{W}$. Curves are for $Q^2 = 0.2, 0.45, 0.85, 1.4$, and $2.4 \GeV^2$ (indicated on the spectra).
        (Left) Ratio $(F_2^{e {}^{56}Fe}/56)/(F_2^{e {}^{12}C}/12)$ versus Bjorken variable $x$ compared to DIS data  \cite{Arneodo:1996rv}.
                }
\label{fig:F2-FeC-em}
\end{figure}
From an experimental point of view, it is also interesting to compare
carbon with iron target nuclei. In the right panel of
Fig.~\ref{fig:F2-FeC-em}, we plot the ratio of structure functions
$(F_2^{e {}^{56}Fe}/56)/(F_2^{e {}^{12}C}/12)$ versus $x$ for several
values of $Q^2$ ranging from $0.2\GeV^2$ to $3.3\GeV^2$. All curves
are shown in the $\xi$ region corresponding to $1.1\GeV<\tilde{W}<2.0
\GeV$.

We stress that there is little 
physical meaning in the fine structure of the curves in the right
panel of Fig.~\ref{fig:F2-FeC-em}. The peaks in the curves,
for example, do not coincide with the resonance peaks. As one can see,
the iron structure functions appear to be very close to the carbon
ones: for each $Q^2$ the ratio of the iron to carbon structure
functions does not deviate more than $5\%$ from the value of $1$. When
averaged, this ratio slightly decreases with increasing $Q^2$, a
behavior which is also exhibited by the DIS data presented in the same
figure. The latter were measured by the NMC Collaboration
\cite{Arneodo:1996rv}, the mean $Q^2$ in the experiment varying from
$20\GeV^2$ for $x \sim 0.1$ to $60 \GeV^2$ for $x \to 1$.


\section{Duality in neutrinoproduction \label{neutrinos}}

In a previous paper \cite{Lalakulich:2006yn} it was demonstrated that
in neutrino reactions quark--hadron duality does not hold for proton
and neutron targets separately. This is a principle feature of
neutrino interactions, stemming from fundamental isospin arguments.
For the charged current reaction $\nu_\mu \, p \to \mu^- \, R^{++}$,
only isospin-3/2 $R^{++}$ resonances are excited, in particular the
$P_{33}(1232)$ resonance. Because of isospin symmetry constraints, the
neutrino-proton structure functions for these resonances are three
times larger than the neutrino-neutron ones.  In neutrino-neutron
scattering, both the isospin-3/2 resonances and the isospin-1/2
resonances contribute to the structure functions. The interplay
between the resonances of different isospins allows for duality to
hold with reasonable accuracy for the average over the proton and
neutron targets. It appears reasonable that one may expect a similar
picture to emerge in neutrino reactions with nuclei.

\begin{figure}[htb]
\begin{minipage}[c]{0.49\textwidth}
        \epsfig{figure=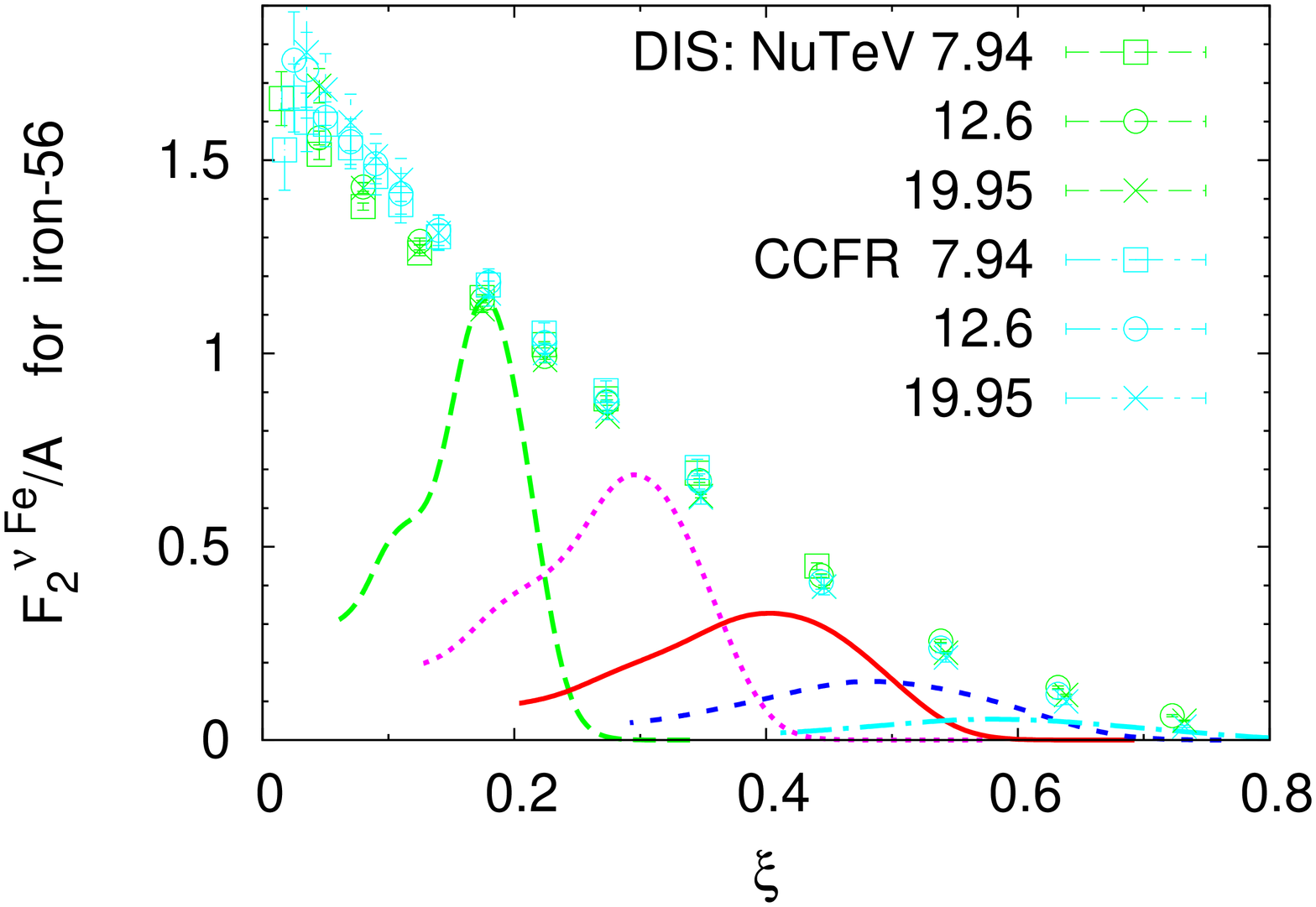,angle=-90,width=\textwidth}
\end{minipage}
\hfill
\begin{minipage}[c]{0.49\textwidth}
        \epsfig{figure=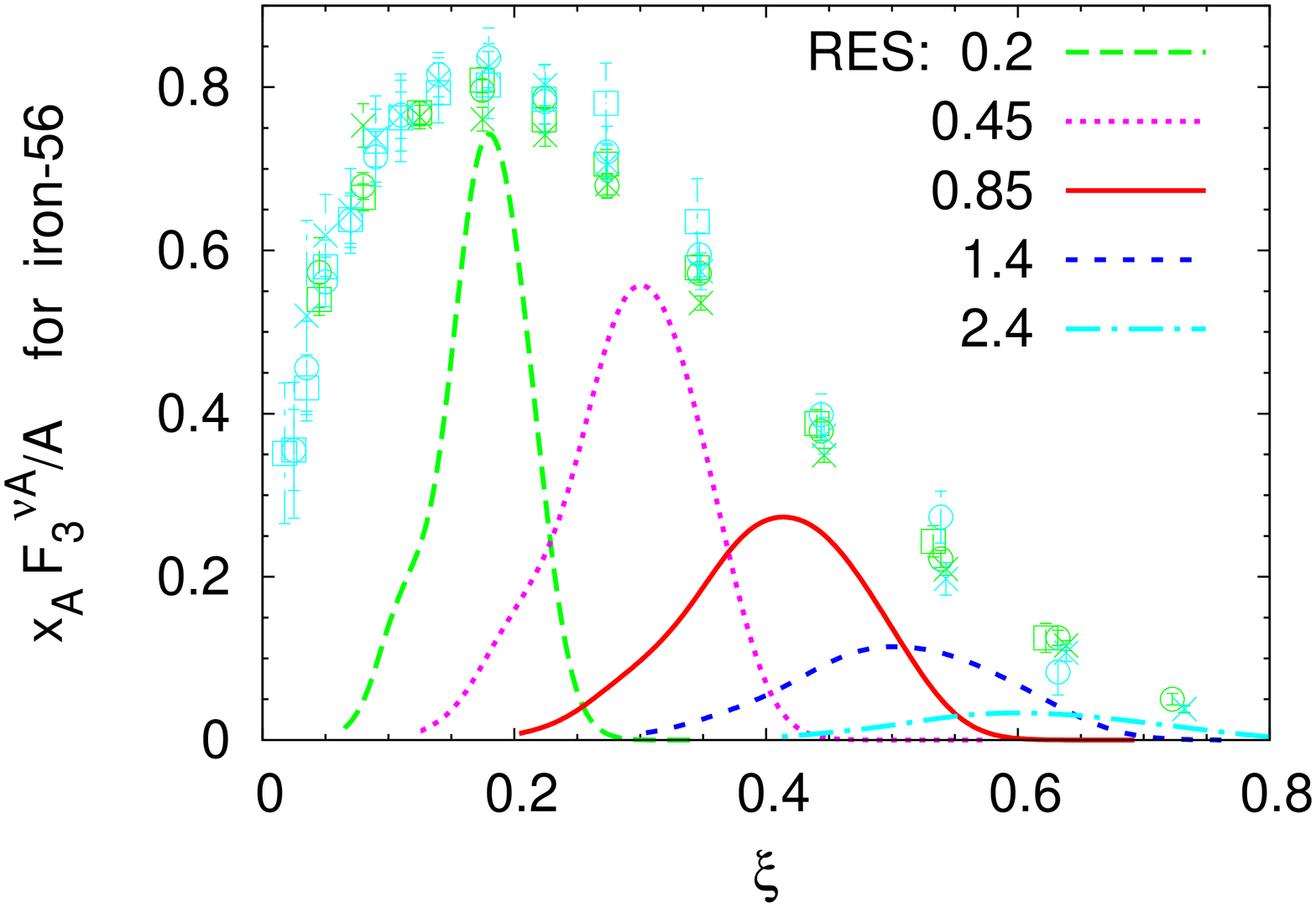,angle=-90,width=\textwidth}
\end{minipage}
\caption{ (color online) The computed resonance curves $F_2^{\nu \,
 {}^{56}\! Fe}/56$ and $x_{Fe} F_3^{\nu \, {}^{56}\! Fe}/56$ as a function of $\xi$,
 for $Q^2 = 0.2, 0.45, 0.85$, $1.4$, and $2.4\; {\mathrm{GeV}}^2$. The
 calculations are compared with the DIS data from
 Refs.~\cite{Seligman:1997mc,Tzanov:2005kr}. The DIS data refer to
 measurements at $Q^2_{DIS}=7.94$, $12.6$ and $19.95 \GeV^2$. 
 }
\label{fig:Fe56-nu}
%
\begin{minipage}[c]{0.49\textwidth}
        \epsfig{figure=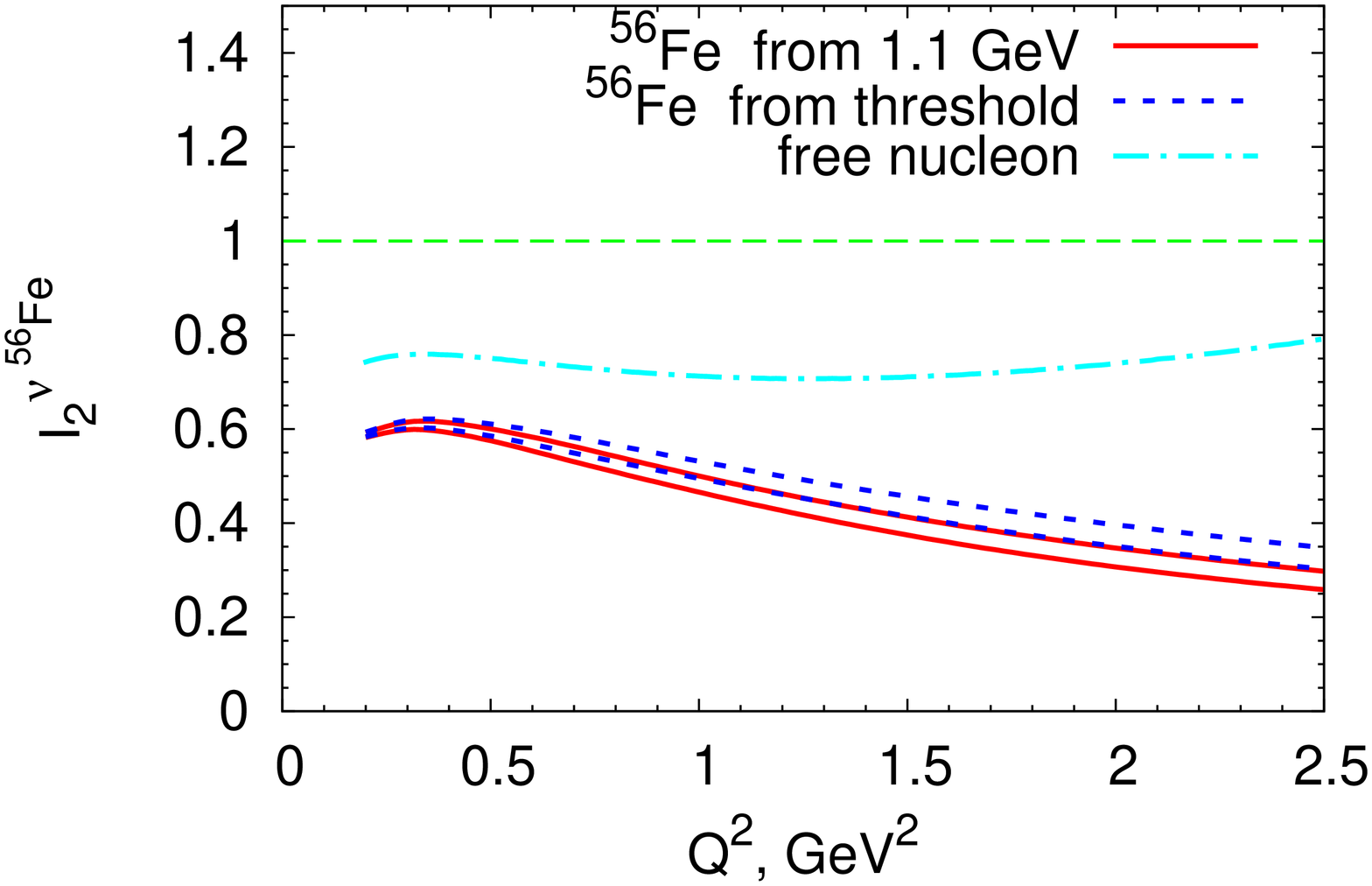,angle=-90,width=\textwidth}
\end{minipage}
\hfill
\begin{minipage}[c]{0.49\textwidth}
        \epsfig{figure=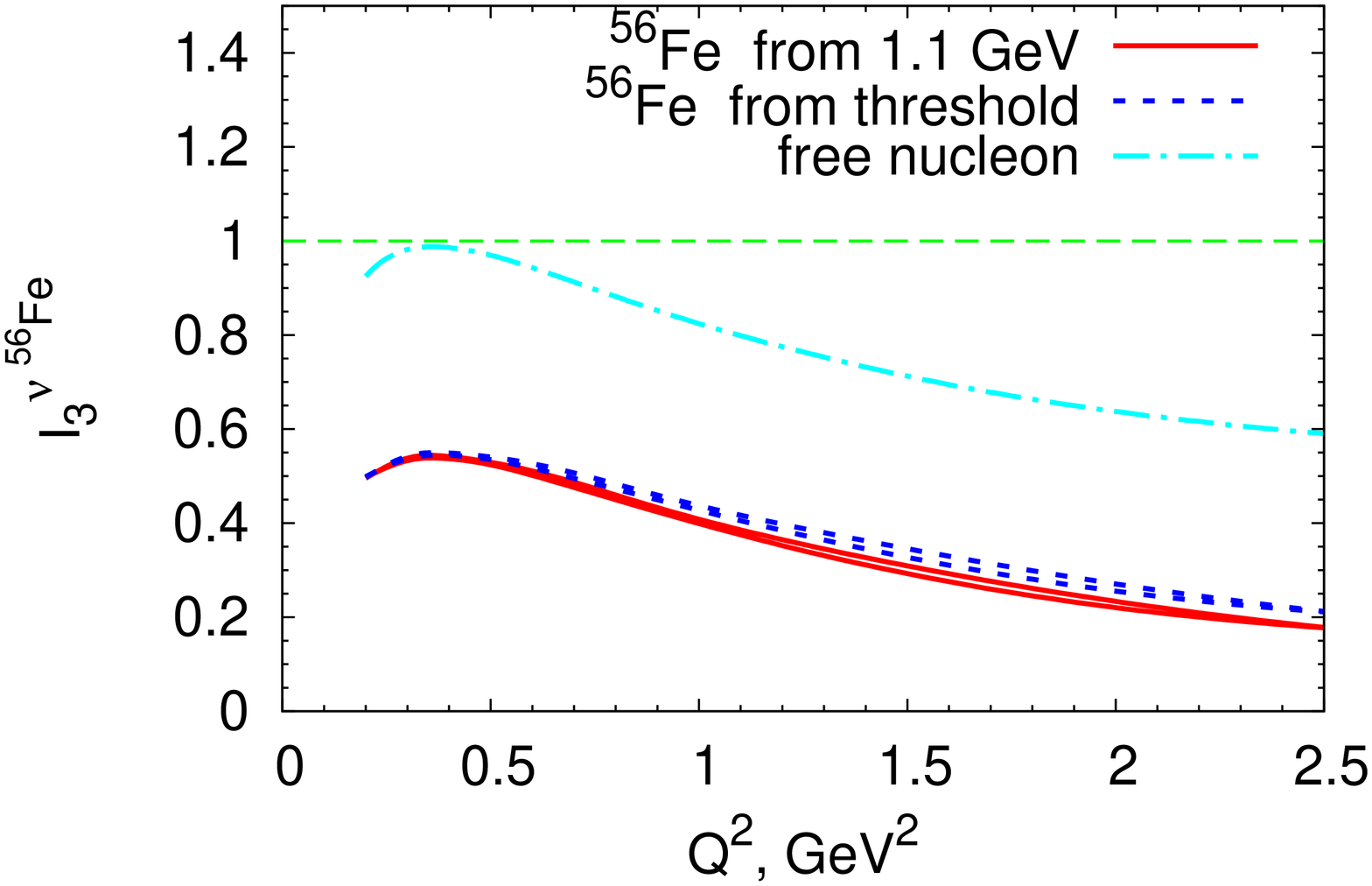,angle=-90,width=\textwidth}
\end{minipage}
\caption{ (color online) Ratios $I_2^{\nu \, {}^{56}Fe}$ and $I_3^{\nu
    \, {}^{56}Fe}$ defined in Eq.~(\ref{eq:Int}) for the free nucleon
    (dash-dotted line) and $^{56}$Fe.  For $^{56}$Fe the results are
    displayed for two choices of the underlimit in the integral:
    $\tilde{W}=1.1 \GeV$ (solid line) and threshold (dotted line).  For
    each of these two choices we have used two sets of DIS data in
    determining the denominator of Eq.~(\ref{eq:Int}).  These sets of
    DIS data are obtained at $Q^2_{DIS}=12.59$ and 19.95 GeV$^2$.  }
\label{fig:I-Fe56-nu}
\end{figure}

The structure functions $F_2^A$ and $x_A F_3^A$ for neutrino--iron
scattering are shown in Fig.~\ref{fig:Fe56-nu}. 
The curves for the
isoscalar free nucleon case is identical to the one presented in
Ref.~\cite{Lalakulich:2006yn} with the ``fast'' fall--off
of the axial form factors for the isospin-1/2 resonances. 

Like for the electron-carbon results of
Fig.~\ref{fig:C12-em}, the resonance structure is hardly
visible. Indeed for each $Q^2$ the computed resonance curves display
one broad peak. The resonance structure functions are compared with
the experimental data in DIS region obtained by the CCFR
\cite{Seligman:1997mc} and NuTeV \cite{Tzanov:2005kr}
collaborations. It appears that the resonance curves slide along the
DIS curve, which indicates global duality. Like for the electron
results discussed in previous section, however, the resonance $F_2^A$
and $x_A F_3^A$ predictions are noticeably lower than the DIS
measurements. 

The ratios $I_2^{\nu \, {}^{56}\! Fe}$ and $I_3^{\nu \, {}^{56}\! Fe}$
defined in Eq.(\ref{eq:Int}) are shown in
Fig.~\ref{fig:I-Fe56-nu}. Our results show, that 1) these ratios are
significantly smaller than 1; 2) they are significantly smaller than
the one for the free nucleon ; 3) $I_2$ is lower than the
corresponding ratio for electroproduction; 4) $I_2$ and $I_3$
slightly decrease with $Q^2$ which is the opposite behavior of what
was observed for electrons.

\begin{figure}[htb]
\begin{minipage}[c]{0.49\textwidth}
        \epsfig{figure=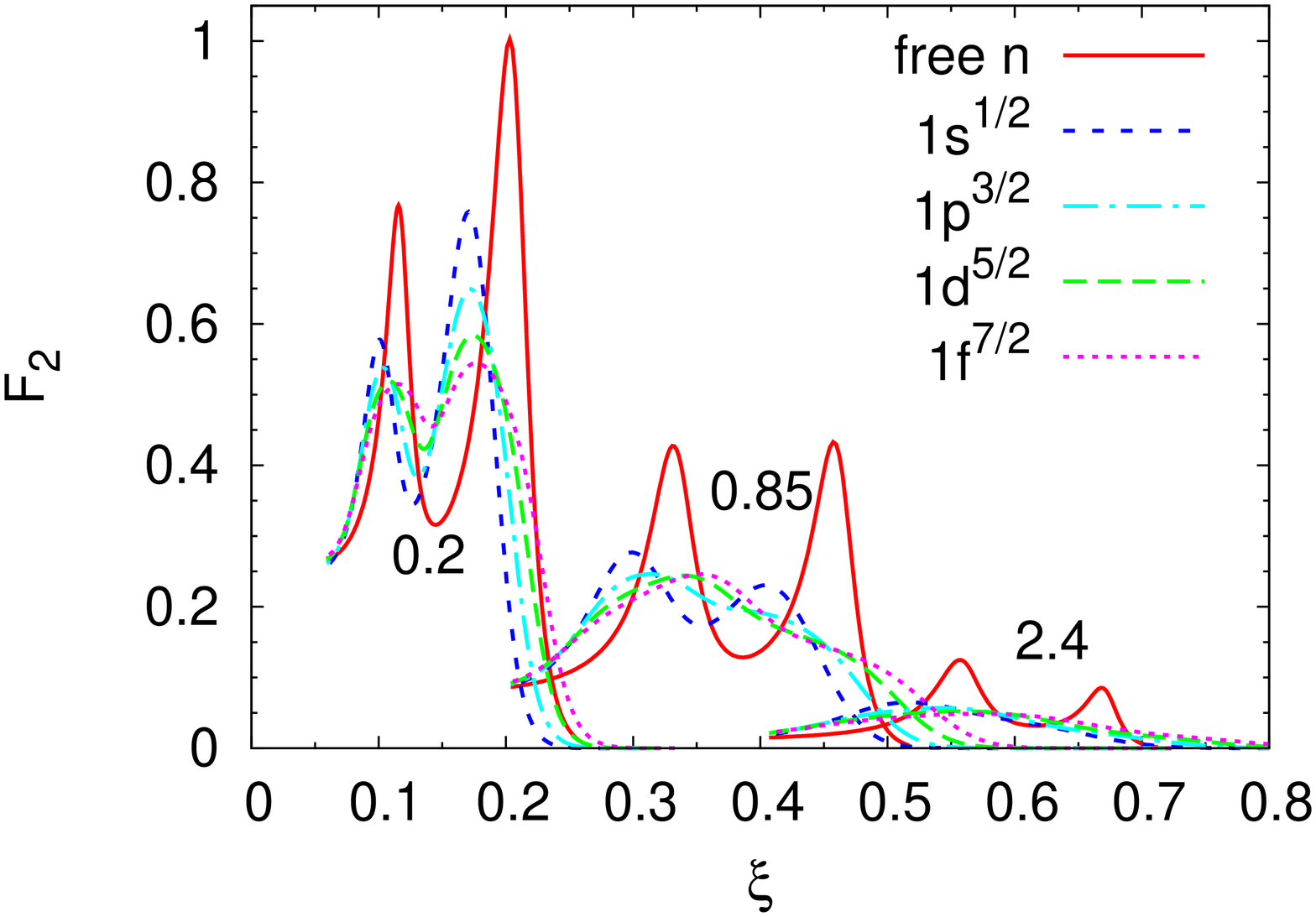,angle=-90,width=\textwidth}
\end{minipage}
\hfill
\begin{minipage}[c]{0.49\textwidth}
        \epsfig{figure=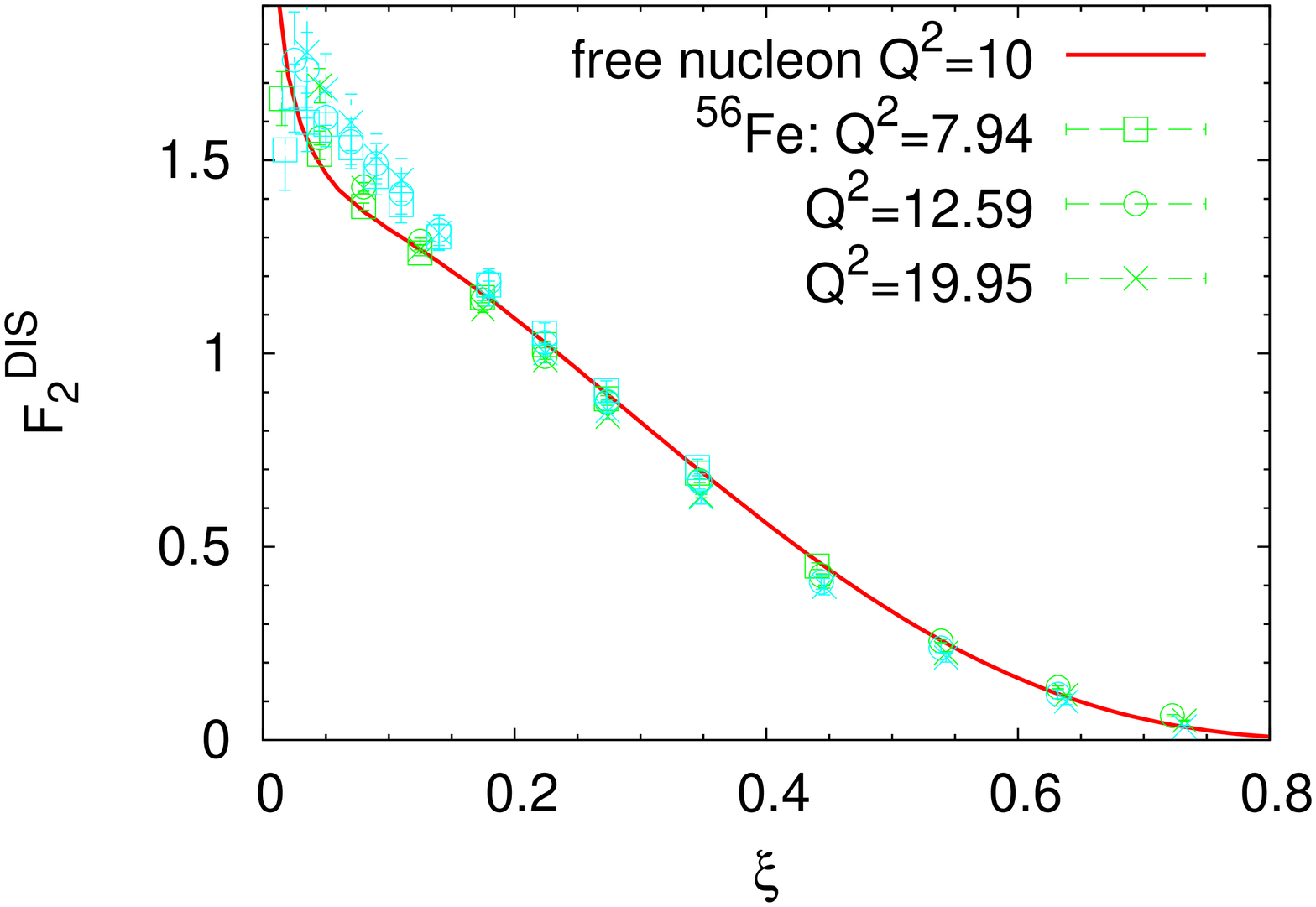,angle=-90,width=\textwidth}
\end{minipage}
\caption{ (color online) 
                        (Left) Weak structure functions $F_2$ in the resonance region for a free neutron (solid curve), for $1s^{1/2}$ (dashed curve), $1p^{3/2}$ (dash--dotted curve), $1d^{5/2}$ (long-dashed curve) and $1f^{7/2}$ (short-dashed curve) neutrons in ${}^{56}Fe$. The three sets of curves correspond to $Q^2=0.2$, $0.85$, and $2.4\GeV^2$.
 			(Right) Weak structure functions $F_2$ in DIS region for a free isoscalar nucleon as obtained via GRV parameterization at $Q^2=10\GeV^2$ (solid curve) and for iron-56 nucleus as measured experimentally \cite{Seligman:1997mc,Tzanov:2005kr} at $Q^2=7.94$, $12.59$ and $19.95 \GeV^2$. 
         }
\label{fig:Fe56-free}
\end{figure}

In an attempt to explain the above observations, we compare the free
isoscalar structure functions with the $^{56}$Fe ones. In the left
panel of Fig.~\ref{fig:Fe56-free} the structure function $F_2^{A}$ for
a neutron in the $1s^{1/2}$, $1p^{3/2}$, $1d^{5/2}$ and $1f^{7/2}$
iron shells are contrasted with the structure function for a free
neutron. For a fixed $Q^2$, the peak at the larger value of the
Nachtmann variable corresponds to the $\Delta$ resonance. The peak at
smaller $\xi$ corresponds to the second resonance region. It is clear
that the nuclear effects reduce the peaks by about $30-50\%$ and shift
them to lower $\xi$ values in comparison with the free nucleon
case. The suppression is most significant for the single-particle
shells close to the Fermi surface.

%

In
close resemblance to what was observed in the discussion of the
electron-nucleus cross sections of previous section, the peculiar
Fermi smearing pattern for each shell introduces additional averaging
when summing over shells.  For a bound proton, the effect of
suppression is nearly the same.  In the right panel of
Fig.~\ref{fig:Fe56-free}, measured DIS structure functions for
$^{56}$Fe at various $Q^2$ are compared to the GRV parameterization
($Q^2=10\GeV^2$) for a free isoscalar nucleon
\cite{Lalakulich:2006yn}. It is obvious that the measured nuclear DIS
structure functions are very similar to the free-nucleon ones. Thus, we
predict a substantial nuclear reduction of the resonance strength,
whereas the data in the DIS region do not point to such a reduction.
This explains the computed low values of the ratios in
Fig.~\ref{fig:I-Fe56-nu}.


We wish to stress that the low values of $I_2^{\nu \, {}^{56}Fe}$ and
$I_3^{\nu \, {}^{56}Fe}$ are not related to the neutron excess. We
remind that the neutron structure functions for the $\Delta$ resonance
are 3 times smaller than the proton ones. The structure
function for isoscalar $^{52}$Fe is only about 5\% larger than for
$^{56}$Fe. This is shown in Fig.~\ref{fig:F23-Fe2Fe-nu} for $F_2$ and
$x_A F_3$. The effect can be easily estimated from  $(26\cdot 3\cdot
f+30\cdot f)/(26\cdot 3\cdot f+26\cdot f)\approx 1.04$, where $f$ is
the neutron structure function in the $\Delta$ region. In the second
resonance region the difference is even smaller.


\begin{figure}[htb]
\begin{minipage}[c]{0.49\textwidth}
        \epsfig{figure=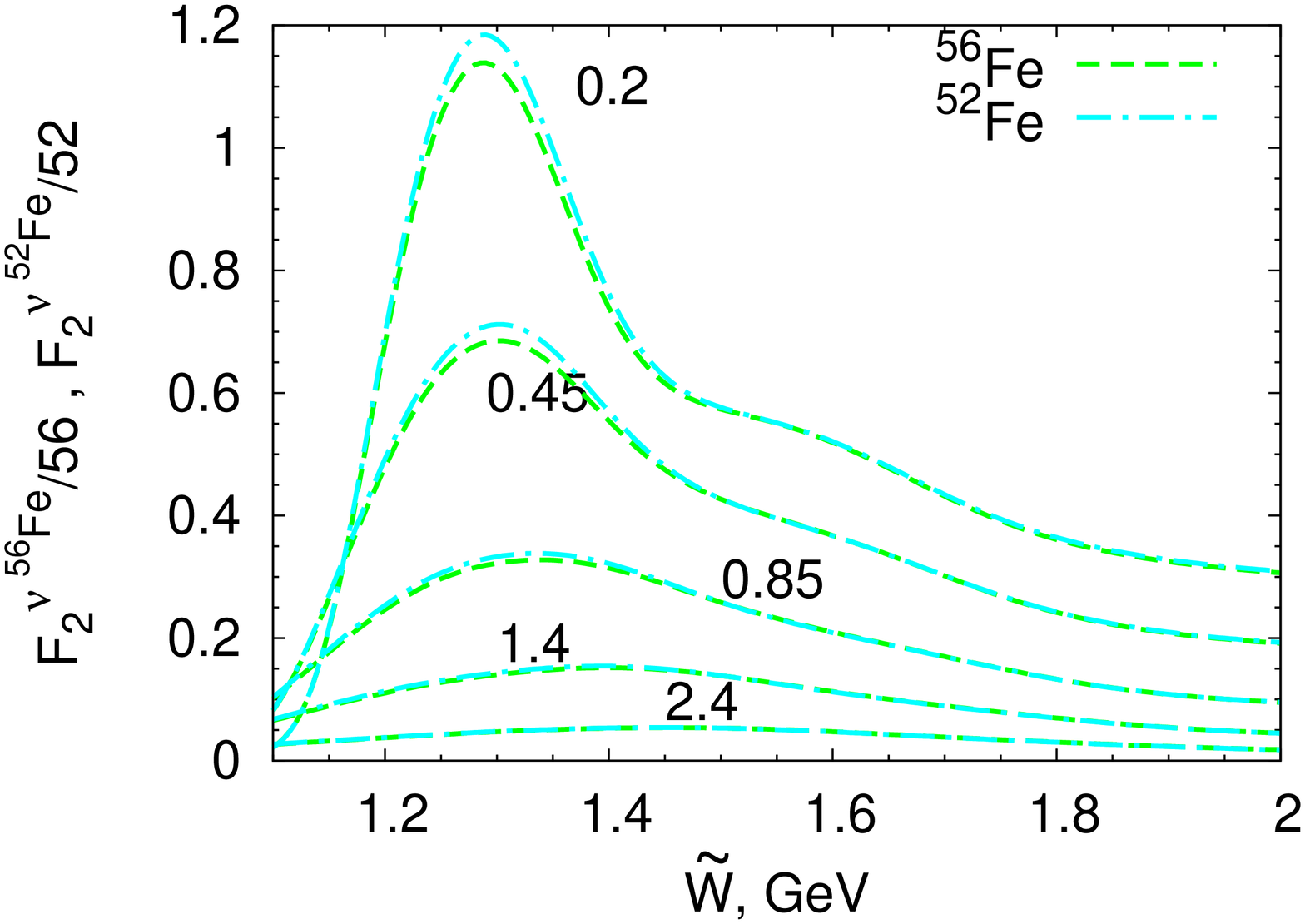,angle=-90,width=\textwidth}
\end{minipage}
\hfill
\begin{minipage}[c]{0.49\textwidth}
          \epsfig{figure=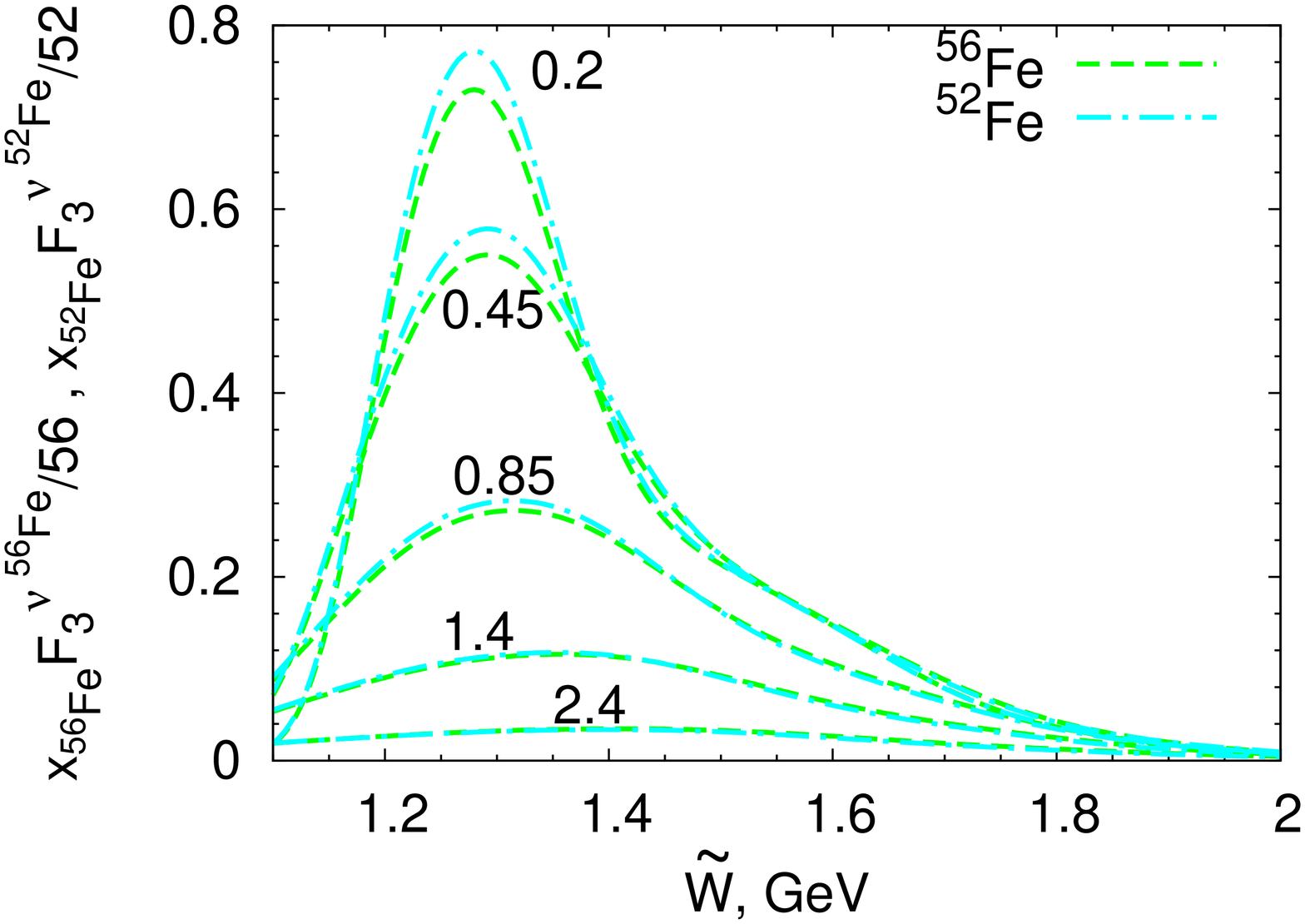,angle=-90,width=\textwidth}
\end{minipage}
        \caption{(color online) Structure functions $F_2^A$ (left) and
        $x_A F_3^A/A$ (right) for neutrinoproduction on iron-52 and
        iron-56 versus $\tilde{W}$. Curves in the resonance region are
        for $Q^2 = 0.2, 0.45, 0.85, 1.4$, and $2.4 \GeV^2$ (indicated
        on the spectra).  
        } 
\label{fig:F23-Fe2Fe-nu}
%
\begin{minipage}[c]{0.49\textwidth}
        \epsfig{figure=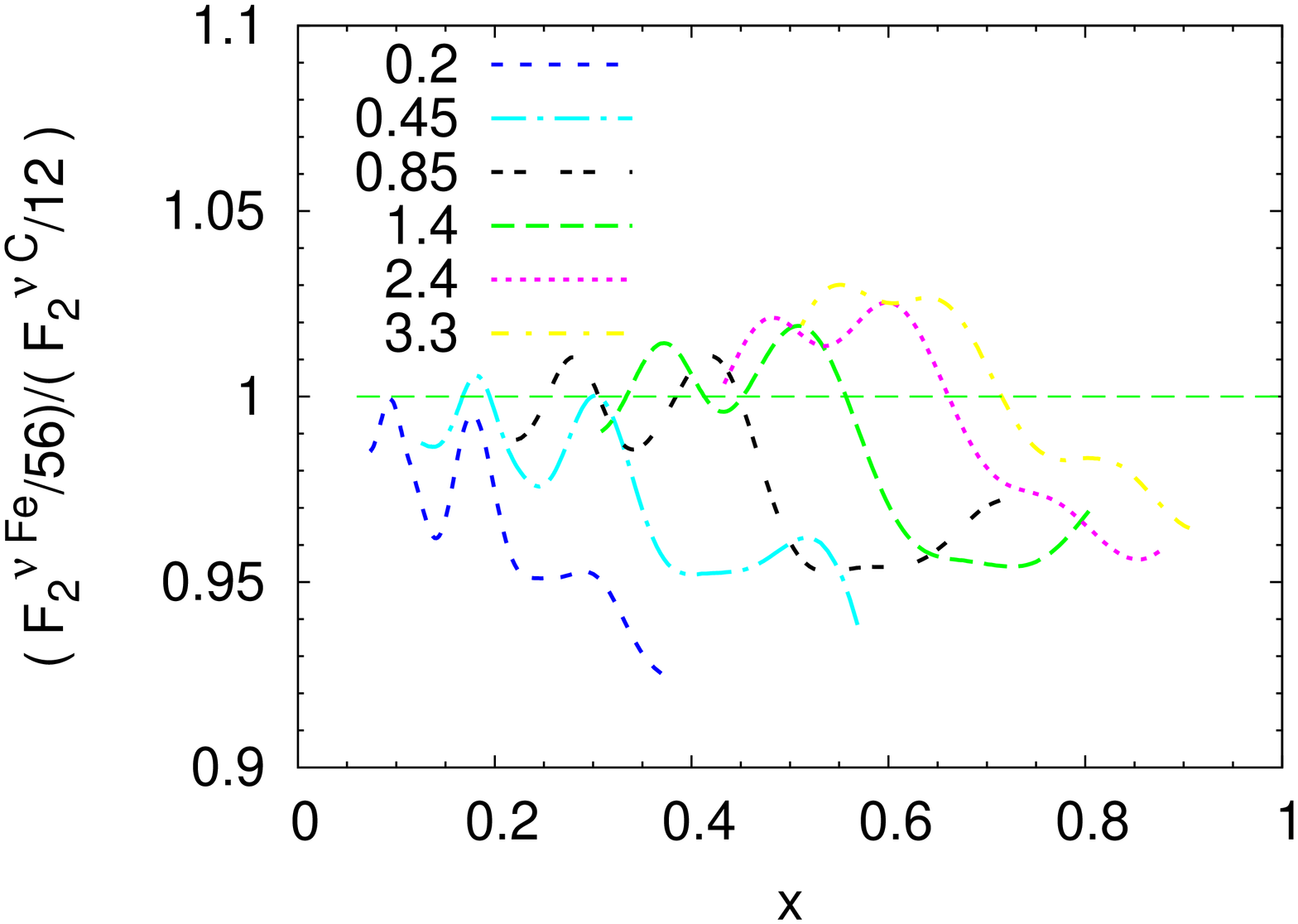,angle=-90,width=\textwidth}
\end{minipage}
\hfill
\begin{minipage}[c]{0.49\textwidth}
          \epsfig{figure=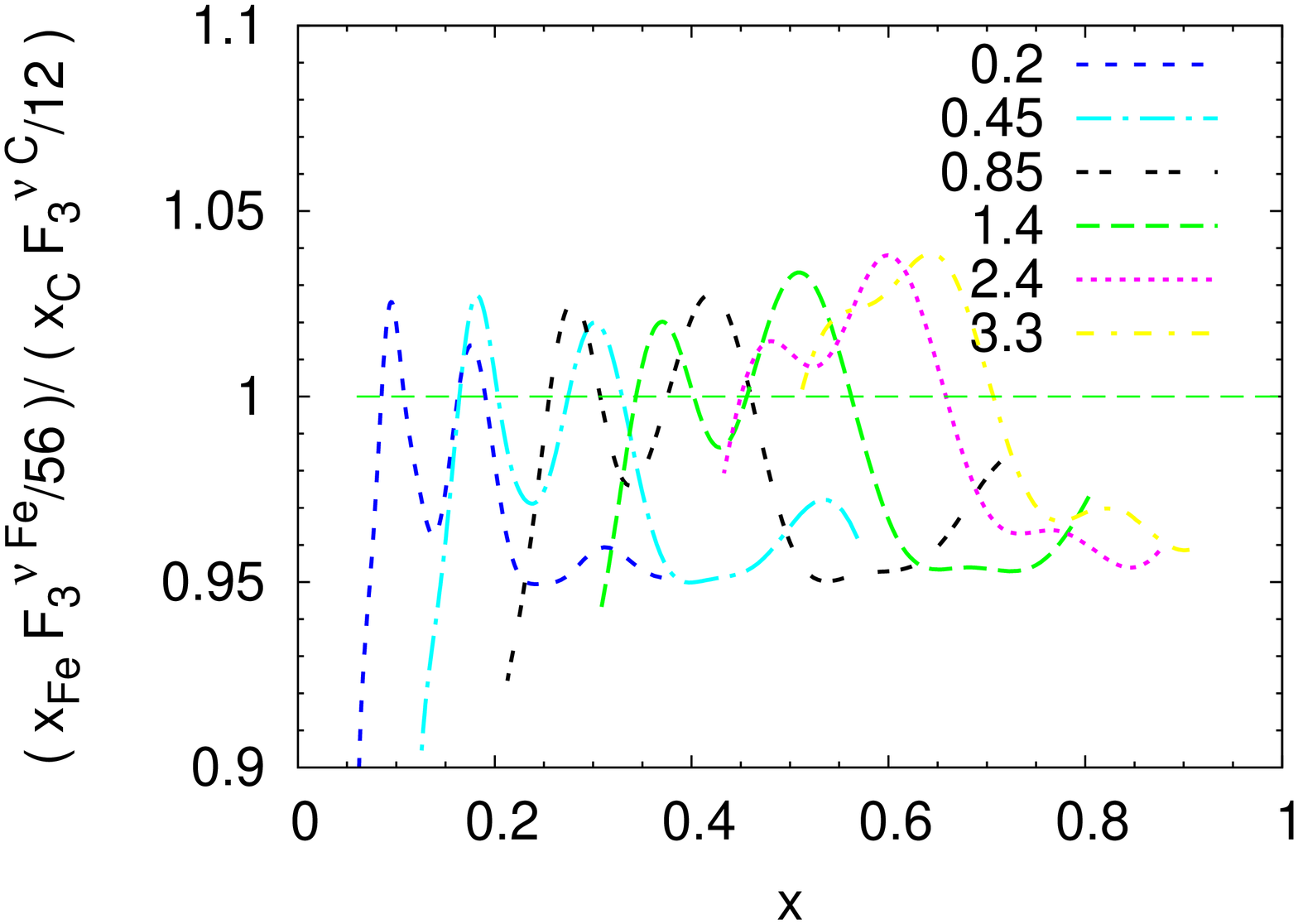,angle=-90,width=\textwidth}
\end{minipage}
        \caption{(color online)  Ratios $(F_2^{\nu \, {}^{56}\! Fe}/56)/(F_2^{\nu \, {}^{12}\! C}/12)$ (left) 
           and $(x_{{}^{56}Fe} F_3^{\nu\, {}^{56}\! Fe/56)}/(x_{{}^{52}\! Fe}F_3^{\nu \, {}^{12}C}/12)$ (right)
                                    versus Bjorken variable $x$ 
                          for $Q^2 =0.2, 0.45, 0.85, 1.4, 2.4$ and $3.3\; {\mathrm{GeV}}^2$
                }
\label{fig:F23-FeC-nu}
\end{figure}

It is also interesting to make a comparison with the carbon
nucleus. The ratios of iron to carbon structure functions $F_2^A$ and
$x_A F_3^A$ versus $x$ are shown in Fig.~\ref{fig:F23-FeC-nu}. For each
$Q^2$ the $\xi$ range corresponds to $1.1<\tilde{W}<2.0 \GeV$.  Like
in the case of electromagnetic reaction, the ratios are close to $1$,
but a bit lower in general and the average is slightly increasing with
$Q^2$. Remark that the peaks in Fig.~\ref{fig:F23-FeC-nu} are not
related to resonances and that  the fluctuations which are of the
order of 5\% can be attributed to subtleties in the shell structure of
the various target nuclei.


\section{Summary \label{summary} }

In view of the current experimental activities, there is great need
for an efficient framework for reliably predicting neutrino--nucleus
cross sections and for a deeper understanding of quark--hadron duality
in nuclei. We performed a phenomenological study of duality in
electron-nucleus and neutrino-nucleus structure functions.  

Using the Dortmund-group model for the production of the {f}{i}rst four
lowest-lying nucleon resonances and using single-particle
wavefunctions from the Hartree approximation to the relativistic
$\sigma \omega$ model, we computed the structure functions $x_A
F_1^A$, $F_2^A$ and
$x_A F_3^A$ in the resonance region for carbon and iron targets and
compared them with the measured DIS ones.  At the same time we
compared the computed resonance structure functions for nuclei with
those for a free nucleon. For quantitative comparisons, we de{f}{i}ned
the ratios $I_{i}(Q^2)$ of integrated resonance to DIS structure
functions. Perfect quark-hadron duality is reached for $I_{i}(Q^2)$
values of unity.  

Summarizing our results, we observe that the computed resonance contribution to
the lepton--nucleus structure functions is qualitatively consistent with the
measured DIS structure functions. This means that global quark--hadron
duality holds for nuclei. The
computed integrated resonance strength, however, is about half of the
measured DIS one.  Contrary to the free nucleon case, where the
ratios $I_{i}(Q^2)$ are at the level of $0.8$, we find for nuclei
$0.6$ for electroproduction and $0.4$ for neutrinoproduction.  This
points towards a scale dependence in the role of the nuclear
effects.  It is obvious that nuclear effects act differently at lower
$Q^2$ (resonance regime) than at higher $Q^2$ (DIS regime).

In our presented analysis we include the resonance contributions and
ignored the role of the background terms. Further investigations
require a theoretical or phenomenological model for the background
contributions in the first and second resonance region.  One could for
example estimate the role of the background contribution to the
$\Delta$--resonance region within the context of the non-linear sigma
model \cite{Hernandez:2007qq}. Extending these or similar models to
higher $W$ values and incorporating them in a model for lepton
reactions with nuclei could be the next step in exploring
quark--hadron duality.

\acknowledgements
The authors acknowledge financial support from the Research Foundation - Flanders (FWO), 
and the Research Council of Ghent University.

\bibliographystyle{apsrev}  
\bibliography{nuclear.bib}

\end{document}